\documentclass[superscriptaddress,nofootinbib,twocolumn,prb]{revtex4-1}

\usepackage{amsmath,amssymb}
\usepackage{graphicx}
\usepackage{color}
\usepackage{hyperref}
\usepackage{url}
\usepackage[utf8]{inputenc}
\usepackage{slashed}
\usepackage{subfigure}
\usepackage{slashed,bbm}
\usepackage{graphics,psfrag,epsfig}
\usepackage{dsfont}
\usepackage{setspace}

\begin{document}

\title{Antiferromagnetism and competing charge instabilities of electrons in strained graphene from Coulomb interactions}

\author{David S\'anchez de la Pe\~na}
\affiliation{Institute for Theoretical Solid State Physics, RWTH Aachen University, 52074 Germany}

\author{Julian Lichtenstein}
\affiliation{Institute for Theoretical Solid State Physics, RWTH Aachen University, 52074 Germany}

\author{Carsten Honerkamp}
\affiliation{Institute for Theoretical Solid State Physics, RWTH Aachen University, 52074 Germany}
\affiliation{JARA-FIT, J\"ulich Aachen Research Alliance - Fundamentals of Future Information Technology}
\affiliation{JARA-HPC, J\"ulich Aachen Research Alliance - High-Performance Computing}

\author{Michael M. Scherer}
\affiliation{Institute for Theoretical Physics, University of Cologne, D-50937 Cologne, Germany}

\begin{abstract}
We study the quantum many-body ground states of electrons on the half-filled honeycomb lattice with short- and long-ranged density-density interactions as a model for graphene. 
To this end, we employ the recently developed truncated-unity functional renormalization group (TU-fRG) approach which allows for a high resolution of the interaction vertex' wavevector dependence.
We connect to previous lattice quantum Monte Carlo (QMC) results which predict a stabilization of the semimetallic phase for realistic \emph{ab initio} interaction parameters and confirm that the application of a finite biaxial strain can induce a quantum phase transition towards an ordered ground state.
In contrast to lattice QMC simulations, the TU-fRG is not limited in the choice of tight-binding and interaction parameters to avoid the occurrence of a sign problem.
Therefore, we also investigate a range of parameters relevant to the realistic graphene material which are not accessible by numerically exact methods.
Although a plethora of charge density waves arises under medium-range interactions, we find the antiferromagnetic spin-density wave to be the prevailing instability for long-range interactions. 
We further explore the impact of an extended tight-binding Hamiltonian with second-nearest neighbor hopping and a finite chemical potential for a more accurate description of the band structure of graphene's $p_z$ electrons.
\end{abstract}

\maketitle

\section{Introduction}


The experimental realization of graphene in 2004\cite{Novoselov2005,Geim2007} has inspired many ideas for a wide range of possible technological applications due to its superior physical properties\cite{castroneto2009,RevModPhys.83.407}, such as its excellent electrical conductivity.
The semi-metallic behavior of graphene's two-dimensional electron gas is protected by the nature of its low-energy excitations, which come in the form of Dirac fermions featuring a linearly vanishing density of states close to the Fermi level.
This has fundamental consequences for the possible effects of many-body interactions\cite{RevModPhys.84.1067}:
For weak electron-electron interactions, the material remains semi-metallic. 
Instead, it requires intermediate to strong interactions to turn it into a Mott insulator or any other ordered many-body ground state\cite{sorella1992,Khveshchenko2001,herbut2006}.
Experimental observations for suspended graphene confirm the stability of the semi-metallic ground state even for very low temperatures\cite{2011NatPh...7..701E,2012NanoL..12.4629M} indicating a subleading role of electronic interactions in graphene.
On the other hand, specific manipulations of the material such as the application of a uniform and isotropic strain have recently been proposed and theoretically found to facilitate the opening of an interaction-induced band gap\cite{PhysRevLett.115.186602}.
This could pave the way towards an even broader range of possible technological applications as, e.g., a graphene transistor. 


The question of whether electronic interactions can induce a metal-insulator transition in an accessible experimental setup can be approached by theoretical methods in two steps: 
(1)~Identification of a suitable model to study interacting electrons in graphene including a determination of model parameters from \emph{ab initio} methods.
(2)~Application of appropriate many-body methods to the model to predict the ground state of the system.


As for (1), the paradigmatic model which is used for the description of the $p_z$ electrons in graphene is composed of a tight-binding Hamiltonian, $H_0$, describing electrons hopping on a honeycomb lattice and an interaction Hamiltonian, $H_1$, which parametrizes the two-body interactions including a short-ranged part and a long-range tail.
For the \emph{ab initio} parameters of the tight-binding Hamiltonian, various works agree on amplitudes of $t\approx 2.7\,\mathrm{eV}$ and $0.02t \lesssim t^\prime \lesssim 0.2t$ for the hopping of an electron to its nearest-neighbor and second-nearest neighbor on the honeycomb lattice, respectively\cite{PhysRevB.66.035412}.
For the determination of the interaction parameters from first principles different methods are available providing different interaction profiles of graphene's $p_z$ electrons\cite{PhysRevLett.106.236805,PhysRevLett.111.036601,PhysRevB.81.085120,PhysRevB.84.085446,PhysRevLett.115.186602}. 
Despite the differences in the details, all methods suggest that the interaction parameters are located in the intermediate coupling regime defining a considerable challenge for many-body methods.

Resulting from considerations of the effects of the different interaction parameters many qualitative studies have revealed a rich ground state manifold depending on the magnitude and ratio of the different local and non-local electron-electron interaction parameters\cite{sorella1992,Khveshchenko2001,herbut2006,PhysRevLett.98.146801,honerkamp2008,herbut2009,raghu2008,PhysRevX.3.031010,PhysRevX.6.011029,PhysRevB.88.245123,grushin2013,daghofer2014,duric2014,PhysRevB.89.195429,PhysRevB.92.085146,PhysRevB.92.155137,volpez2016,sanchez2016}.
Possible ground states include an antiferromagnetic spin-density wave state, different commensurate and incommensurate charge density wave states, a Kekul\'e dimerization pattern and more.
More recently, numerically exact methods, i.e. quantum Monte Carlo (QMC) simulations have become available which can explore a range of realistic parameters for the graphene model\cite{PhysRevLett.111.056801,PhysRevB.89.195429,Buividovich:2016tgo}.
These works confirm the experimentally found semi-metallic behavior of the material. 
It was further suggested that a biaxial strain of about 15\% can turn graphene into an antiferromagnetic Mott insulator\cite{PhysRevLett.115.186602} at least when the Thomas-Fermi method for the determination of the interaction profile\cite{PhysRevB.84.085446} is assumed.
On the other hand, the {\it ab initio} interaction profile suggested by the constrained random phase approximation (cRPA)\cite{PhysRevLett.106.236805} did not indicate a semi-metal-insulator transition up to 18\% strain\cite{PhysRevLett.115.186602}.
It may be noted, however, that QMC methods are limited by the choice of band structure and interaction parameters\cite{Buividovich:2016tgo,PhysRevLett.115.186602}, i.e. to avoid the occurrence of a sign problem, the long-range tail of the electronic interaction profile has to decrease fast enough.
Therefore, a third option for the interaction parameters from the Pariser-Parr-Pople model\cite{PhysRevB.81.085120} could not be investigated in Ref.~\onlinecite{PhysRevLett.115.186602}.
Also, band structure parameters other than the nearest-neighbor hopping $t$ have to be neglected within QMC simulations.
This introduces a bias to the range of available results when aiming at the description of realistic graphene models.
More specifically, the limitation of the interaction profile to a long-range behavior that pronounces the local part of the interaction leads to a preference of the antiferromagnetic ground state.
In fact, the antiferromagnetic ground state is the only ordered state that has been accessed by QMC simulations with one exception: In Ref.~\onlinecite{Buividovich:2016tgo} a model with onsite interaction $U$ and nearest-neighbor interaction $V_1$ was studied and indications for a competition between spin- and charge-density wave order have been found for specific choices of parameters giving rise to a multicritical point in parameter space, cf. also Refs.~\onlinecite{PhysRevB.92.035429,PhysRevB.93.125119}.


In this paper, we overcome the limitations in the choice of interaction profiles by employing a recently developed implementation of the functional renormalization group\cite{Wetterich:1992yh} approach for correlated fermions\cite{doi:10.1143/PTP.105.1,Metzner2012} which allows for a high resolution of the interaction vertex' wavevector dependence -- the truncated-unity fRG (TU-fRG)\cite{Lichtenstein2016,sanchez2016} -- making use of high-performance computing facilities\cite{2016arXiv161009991L}.
In particular, this allows to explore a large set of band structure parameters, e.g., a second-nearest neighbor hopping term and a chemical potential as well as an extended range of realistic interaction parameters.
Recent TU-fRG calculations for an explorative set of short-ranged interaction parameters already suggest that the semimetallic nature of graphene's groundstate is not due to interaction terms that are too weak to induce an ordered state, but rather because of a complex interplay between different competing instabilities which leads to an effective frustration\cite{sanchez2016}.
Moreover, it has been found that the leading instability is not necessarily an antiferromagnetic spin density wave state, but can also be an incommensurate charge density wave and other charge modulated states\cite{sanchez2016}.
We note that the fRG is not a method which provides numerically exact results, however, numerically exact methods have a much narrower scope. In the situations accesible to exact methods, a systematic comparison with fRG results provides confidence for the method's application to other regimes. This provides important insights on the real material, allowing to go beyond the statements that are possible within a single theoretical method alone.

In this work, we employ the TU-fRG to facilitate an unbiased approach to identify the leading instability of electrons on the honeycomb lattice with realistic band structure parameters and a long-range interaction tail provided by \emph{ab initio} approaches.
As a particular strength of the (TU-)fRG approach in this context, we emphasize that it does take into account the fermionic fluctuations in an unbiased way.
Furthermore, the TU-fRG is not bound to a sufficiently fast decay of the (partially screened) Coulomb tail and provides a sufficient wavevector resolution to resolve the long-range tail.
In particular, it can explore the effect of arbitrary ratios of short-ranged (non-local) interaction terms which are known to trigger very different types of order.
This is a clear advantage to the numerically exact QMC methods which have a sign problem if the Coulomb tail does not decay sufficiently fast and are therefore 'biased' towards antiferromagnetic order.
So, while our results will not give quantitative estimates for gaps or transition temperatures, we will be able to resolve the qualitative effects and competing orders that are induced by an extended range of realistic interaction profiles.

A broader scope of the insights obtained in this work is given by the more general set of low-dimensional $sp$-electron systems of adatoms on semiconductor surfaces such as Si(111):X with X=C, Si, Sn, Pb which exhibit both strong local and non-local Coulomb interactions. In these surface systems, e.g., for Si:X, Mott transitions have been observed, cf. Ref.~\onlinecite{PhysRevLett.110.166401}.

The paper is organized as follows: In Sec.~\ref{sec:model}, we introduce the model Hamiltonian including the tight-binding part and the interaction part. We present possible choices for its parameters as suggested by various {\it ab initio} methods and discuss the effect of finite biaxial strain. Sec.~\ref{sec:method} shortly introduces the functional renormalization group approach with a focus on correlated fermion systems. More specifically, we also discuss the TU-fRG truncation scheme, which is employed here to facilitate calculations with high wave-vector resolution. Technical details on this scheme are presented in App.~\ref{sec:tufrg}. In Sec.~\ref{sec:results}, we show our results by first discussing the impact of a finite second-nearest neighbor hopping amplitude, then by systematically extending the range of interaction terms  step by step and finally by including a large range of non-local interactions suggested by the different {\it ab initio} methods. Conclusions are drawn in Sec.~\ref{sec:conc} and further technical details are discussed in the appendix.

\section{Model and Parameters}\label{sec:model}

\begin{figure}[t!]
\includegraphics[height=0.51\columnwidth]{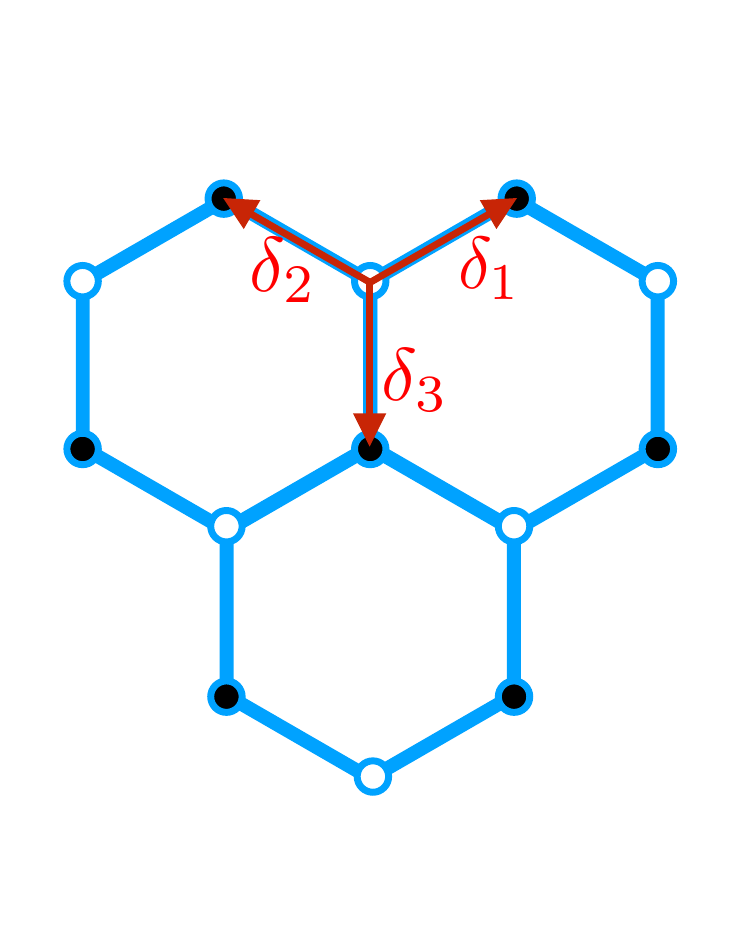}
\includegraphics[height=0.51\columnwidth]{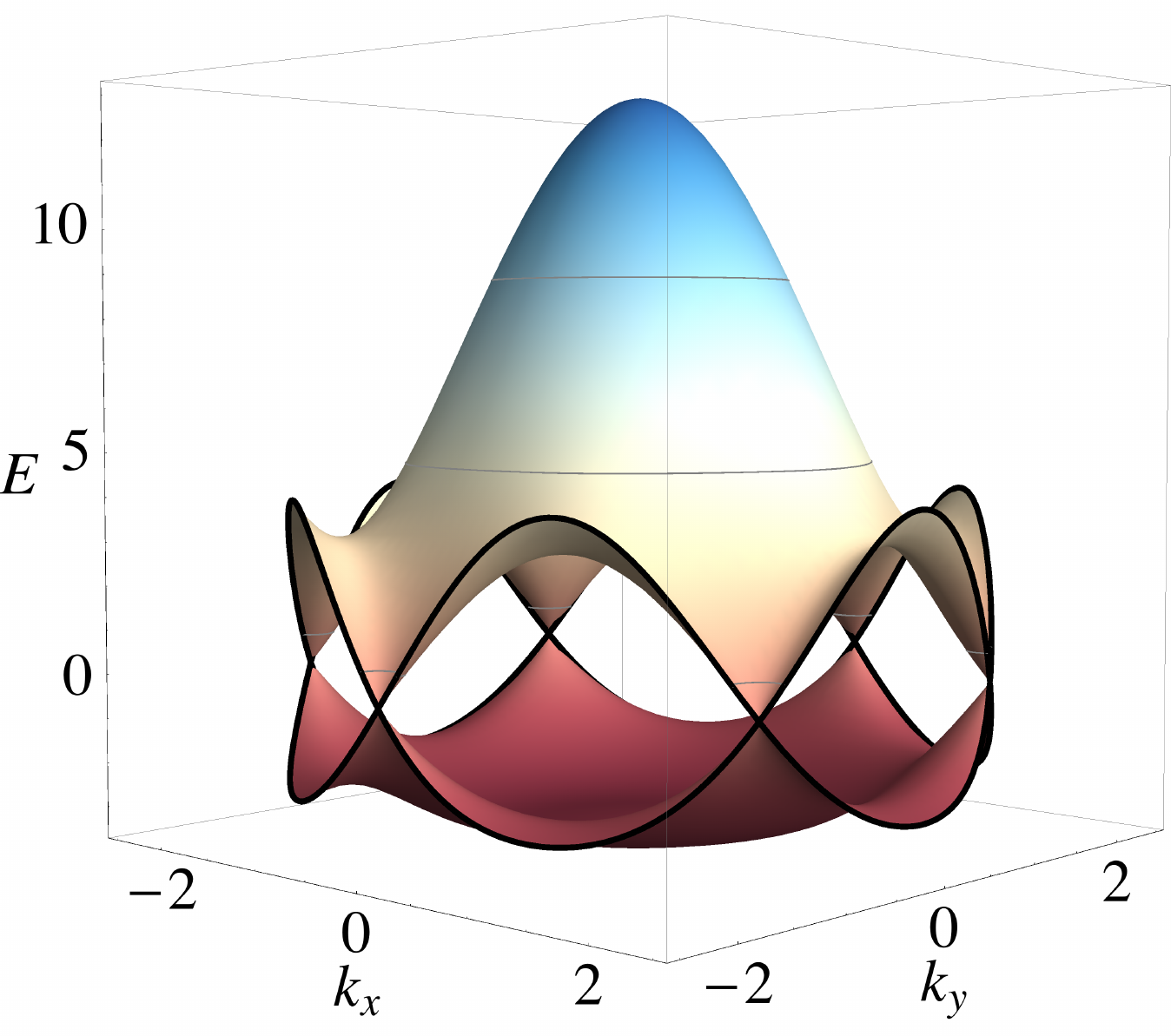}
\caption{Left panel: Real space lattice structure. Right panel: band energy dispersion for the tight-binding parameters $t=2.7\,\mathrm{eV}$ and $t^\prime=0.2\,t$ with adjusted chemical potential. Energy units are given in eV.}
\label{fig:model}
\end{figure}

To model the interacting $p_z$ electrons on graphene's half-filled honeycomb lattice, we consider a tight-binding model for spin-1/2 fermions enhanced by density-density interaction terms representing the long-ranged Coulomb interaction.
Therefore, the Hamiltonian has a single-particle hopping term~$H_0$ and an interaction term~$H_1$,
\begin{equation}\label{modelHamiltonian}
 H=H_0+H_1\,,
\end{equation}
to be specified in the following.
$H_0$ is the tight-binding part
\begin{align}\label{eq:sp}
 H_0=&-t \sum_{\langle i,j \rangle,\sigma} \left( c^{\dagger}_{i,A,\sigma} c_{j,B,\sigma} +\text{H.c.} \right)\nonumber\\
 	&-t^\prime \!\! \sum_{\langle\langle i,j \rangle\rangle,\sigma}\left( c^{\dagger}_{i,A,\sigma} c_{j,A,\sigma} +c^{\dagger}_{i,B,\sigma} c_{j,B,\sigma} +\text{H.c.} \right)\nonumber\\
	&-\mu \sum_{i,\sigma}\left( c^{\dagger}_{i,A,\sigma} c_{i,A,\sigma} +c^{\dagger}_{i,B,\sigma} c_{i,B,\sigma}\right)\,,
\end{align}
with nearest-neighbor hopping amplitude $t$, second-nearest neighbor hopping $t^\prime$ and chemical potential $\mu$. The nearest-neighbors are given by the position vectors $\vec{\delta}_1, \vec{\delta}_2, \vec{\delta}_3$ of the hexagonal lattice, depicted in Fig.~\ref{fig:model}, which has a two-atomic basis with sublattice index $o \in \{A,B\}$. We will interchangeably denote $o$ as sublattice or orbital degrees of freedom, not to be confused with different orbitals within a single atomic site. The Carbon-Carbon distance is normalized to unity, i.e. $|\vec{\delta}_i|=a=1$.
Moreover, $c^{(\dagger)}_{i,o,\sigma}$ annihilates (creates) an electron at site $i$ in sublattice $o$ with spin $\sigma \in \{\uparrow,\downarrow\}$.

This tight-binding model gives graphene's characteristic valence and conduction bands which touch linearly at the two inequivalent corner points of the Brillouin zone (BZ), i.e. the $K, K^\prime$ or Dirac points as described by the energy dispersion $E_\pm=\pm t\sqrt{3+d(\vec{k})}-t^\prime d(\vec{k})-\mu$ with $d(\vec{k})=2\cos\left(\sqrt{3}k_y\right)+4\cos\left(\frac{\sqrt{3}}{2}k_y\right)\cos\left(\frac{3}{2}k_x\right)$.
Close to the Dirac points the energy dispersion can be approximated by $E_\pm\approx 3t^\prime \pm \frac{3t}{2}|\vec{q}|$, i.e. to put the Fermi-level to the Dirac points we have to adjust the chemical potential to $\mu=3t^\prime$.
For the \emph{ab initio} parameters of the tight-binding Hamiltonian, suggested amplitudes are $t\approx 2.7\,\mathrm{eV}$ and $0.02t \lesssim t^\prime \lesssim 0.2t$, cf. Ref~\onlinecite{PhysRevB.66.035412}.

The interaction part $H_1$ from the Coulomb interaction of the electrons is parametrized by local and non-local density-density contributions
\begin{align}\label{Hint}
 H_1 = \, &U \sum_{i,o} n_{i,o,\uparrow} n_{i,o,\downarrow} \nonumber\\
 &+ \sum_{\substack{i\neq j, \, o,o^\prime \\ \sigma,\sigma'}} \frac{U_{i,j}^{o,o^\prime}}{2} n_{i,o,\sigma} n_{j,o^\prime,\sigma'}\,,
\end{align}
where $n_{i,o,\sigma}=c^{\dagger}_{i,o,\sigma} c_{i,o,\sigma}$ represents the electron density operator, and interaction coefficients read
\begin{equation}
 U_{i,j}^{o,o^\prime} = U_{i,j} \begin{cases}
			  \delta_{o,o^\prime} &\text{for intra-orbital $(i,j)$ pairs} \\
			  1 - \delta_{o,o^\prime} &\text{for inter-orbital $(i,j)$ pairs}
			\end{cases}
\end{equation}

Different kinds of ordered states occur when the individual interaction parameters exceed critical values.
Sizable onsite interactions $U>0$ trigger a phase transition towards an antiferromagnetic spin-density wave (SDW) state. 
Each $n^{\text{th}}$ nearest-neighbor repulsion term $U_{i,i+n} = V_n$ supports a different ordering transition towards charge order, with the nearest-neighbor term $V_1$ triggering the conventional charge-density wave (CDW).
%

\subsection{Modification of hopping amplitudes from strain}

The hopping amplitudes in the tight-binding Hamiltonian in Eq.~\eqref{eq:sp} are subject to modifications upon lattice distortions as a result of the change in wave-function overlap. 
For the \emph{ab initio} model parameters from the constrained random phase approximation of Ref.~\onlinecite{PhysRevLett.106.236805}, $t$ has a linear decay vs.~strain $\eta$. To model the effect of strain on other choices of \emph{ab initio} model parameters, where direct results are not available, we assume an exponential decay of the hopping amplitudes following the empirical relation\cite{PhysRevB.80.045401,PhysRevB.88.115428}
\begin{align}
	t_{ii^\prime}=t_0 e^{-\beta\left(\frac{|\vec{\delta}_{ii^\prime}|}{a}-1\right)}\,,
\end{align}
where $a$ is the unstrained lattice constant which we have set to $a=1$ and $t_0$ is the unstrained nearest-neighbor hopping amplitude. 
The material-dependent factor $\beta$ is estimated to $\beta=3.37$ for graphene and $\vec{\delta}_{ii^\prime}$ is the vector connecting sites $i$ and $i^\prime$. For the unstrained second-nearest-neighbor hopping, this formula provides an numerical value of $t^\prime=t_0 \exp(-3.37(\sqrt{3}-1))\approx 0.085t$ which is located in the estimated range.

A finite and uniform strain $\eta$ can be included using a strategy suggested in Ref.~\onlinecite{PhysRevLett.115.186602} by the replacement $|\vec{\delta}_{ii^\prime}|=r\rightarrow (1+\eta)r$ with strain parameter $\eta$.
Then, the strained hopping amplitudes are given by
\begin{align}\label{eq:hopstrain}
	t(r,\eta)=t_0 e^{-3.37\cdot\left[(1+\eta)r-1\right]}\,,
\end{align}
and $r$ has to be evaluated at the equilibrium positions of the neighboring sites, i.e. $r=1$ for the nearest neighbor and $r=\sqrt{3}$ for the second-nearest neighbor. This gives a strain-dependence of the nearest-neighbor hopping of $t(\eta)=t_0\exp(-3.37\eta)$ and for the second-nearest neighbor hopping $t^\prime(\eta)=t_0\exp(-3.37((1+\eta)\sqrt{3}-1))$.

\subsection{\emph{Ab initio} interaction parameters}\label{sec:int_param}

For the determination of the interaction parameters for $p_z$ electrons in graphene from first principles various methods are available. 
In the context of biaxially strained graphene and its effect on the quantum many-body ground states, three of these methods have been explored in Ref.~\onlinecite{PhysRevLett.115.186602} for the case of graphene: 
The Thomas-Fermi (TF) method\cite{PhysRevB.84.085446}, the constrained random phase approximation (cRPA)\cite{PhysRevLett.106.236805} and the quantum-chemistry-Pariser-Parr-Pople (QC-PPP) method\cite{PhysRevB.81.085120}. 
In this work, we disregard the TF method, which shows the strongest decay in the interaction parameters when going to larger distances. 
Therefore, this method can be considered to be well-covered by the QMC simulations.
Instead, here we concentrate on the cRPA and the QC-PPP methods which have stronger non-local short-ranged interactions.
In particular, due to a sign problem, it was not possible to study the interaction profile suggested by the QC-PPP method and we fill this gap, here.

\subsubsection{Constrained random phase approximation}

In the cRPA the effective interaction profile for graphene's $p_z$ electrons is described by the formula
\begin{align}
	V(r)=\frac{\bar V(r)}{1-\bar V(r)P(r)}\,,
\end{align}
where $\bar V(r)$ is the bare Coulomb potential and $P(r)$ is a polarization function.
Explicit values for onsite interaction $U$, nearest-neighbor interaction $V$ and the nearest neighbor hopping $t$ for unstrained and strained graphene have been calculated in Ref.~\onlinecite{PhysRevLett.106.236805}, exhibiting a linear dependence of $U,V,t$ on strain.
We directly take the values therein, available till the fourth-nearest neighbor, as the input for our calculations. The longer-ranged part of the Coulomb-tail is affected by the surrounding electrons leading to a modified dielectric constant, i.e. $1/r \rightarrow 1/[r(1+\pi\, r_s/2)]$ where $r_s=e^2/(\kappa\hbar v_F)$ is the Wigner-Seitz radius of monolayer graphene which depends on the fermi velocity $v_F=\frac{\sqrt{3}}{2}t\,a$,  with $a=a_0(1+\eta)$. Alternatively, we parametrize the Coulomb-tail with an artificial dielectric constant $\epsilon$, i.e. $1/r \rightarrow 1/(\epsilon\, r)$, which is extrapolated from the available short range terms.

We note that in the limit $r\to\infty$ the Coulomb potential approaches $1/r$ again, i.e. $\epsilon \rightarrow 1$, as the two-dimensional fermion degrees of freedom cannot modify the three-dimensional Coulomb potential. 
Here, we do not take into account this latter effect.
For better comparison, the cRPA values of terms other than $U,V$ under strain are taken to be the same values as in Ref.~\onlinecite{PhysRevLett.115.186602}.

\subsubsection{Ohno interpolation formula}

In the context of biaxially strained graphene it was suggested in Ref.~\onlinecite{PhysRevLett.106.236805}, that the Coulomb interaction can be modeled by the Ohno interpolation formula\cite{Ohno1964} 
\begin{align}\label{eq:ohno}
	V(r_{ij},\epsilon)=\frac{U}{\sqrt{1+\left(\epsilon\, \frac{U}{e^2} r_{ij}\right)^2}}\,,
\end{align}
where $V(0)=U$ and $\epsilon$ is a variable screening and for large distances $r\rightarrow\infty$ approaches $V(r)\to e^2/(\epsilon\, r)$.
The screening parameter $\epsilon$ can generally be tuned in the intervall $\epsilon \in [0,\infty)$, where $\epsilon\to\infty$ results in a purely local onsite interaction $V(r_{ij},\infty)=U\,\delta_{ij}$. Furthermore, $\epsilon=0$ is the limit of a constant (non-local) interaction $V(r_{ij},0)=U$ and $\epsilon=1$ represents the case of benzene\cite{Bursill1998305}.
Ref.~\onlinecite{PhysRevLett.106.236805} argues that employing the values for the interaction parameters $U$ and $V_1$ as given for phenalenyl (3H -- C${}_{13}$H${}_{9}$) molecule from the quantum-chemistry-Pariser-Parr-Pople (QC-PPP) method provide an upper bound for the Hubbard $U$ and the interaction potential $V(r)$, see Ref.~\onlinecite{PhysRevB.81.085120}.
The transformation matrix for the interaction profile as given by the QC-PPP method is not positive definite, therefore it was not accessible to the QMC methods promoted in Ref.~\onlinecite{PhysRevB.81.085120}.
We explicitly study this type of interaction profile and variations of it taking account for the fact that the interaction parameters are only known approximately.

A finite strain $\eta$ can be included employing the strategy suggested in Ref.~\onlinecite{PhysRevLett.115.186602}: Replace $r\rightarrow (1+\eta)r$ in $V(r)$ and use $t\rightarrow t_0 e^{-3.37\eta}$.
The QC-PPP method is designed to describe small system sizes and larger systems are expected to show stronger screening and therefore a smaller $V(r)$.
We therefore interpret these parameters as providing an upper limit for a realistic choice of the interaction profile and note that extrapolation to larger systems has to be interpreted cautiously.

\section{Method}\label{sec:method}

\begin{figure}[t!]
  \centering
 \includegraphics[height=0.51\columnwidth]{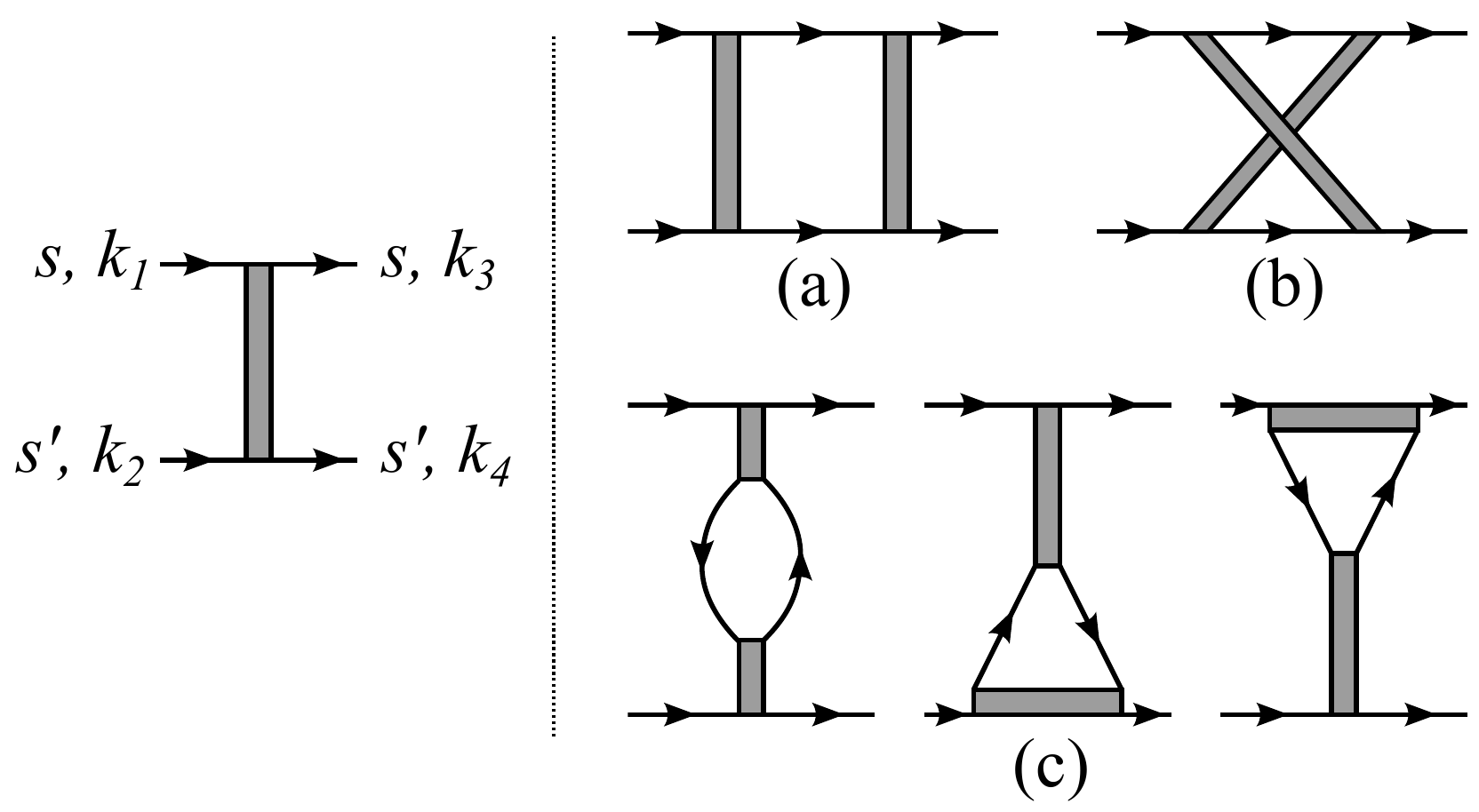}
  \caption{Left panel: Interaction vertex with conventions for wave-vectors and spin projections. Right panel: Diagrammatic representation of the right-hand side of Eq.~\eqref{eq:flow}, consisting of one-loop (a) particle-particle, (b) crossed particle-hole and (c) direct particle-hole contributions.}
  \label{fig:diag}
\end{figure}

To study the quantum many-body instabilities of interacting fermion systems, we employ the functional Renormalization Group approach\cite{Wetterich:1992yh,doi:10.1143/PTP.105.1,Metzner2012} which describes the evolution of the one-particle irreducible (1PI) vertex functions upon integrating out high-energy fermionic modes. In a standard level-two truncation, the interacting system is described by an effective two-particle interaction term which is proportional to
\begin{align}
V_\Omega^{b_1,b_2,b_3,b_4}(\mathbf{k}_1,\mathbf{k}_2,\mathbf{k}_3)c^\dagger_{b_4,\mathbf{k}_4, \sigma}c^\dagger_{b_3,\mathbf{k}_3, \sigma^\prime}c_{b_2,\mathbf{k}_2, \sigma}c_{b_1,\mathbf{k}_1, \sigma^\prime}\,,
 \end{align}
depending on four band indices and three momenta in the presence of translational and the spin-SU(2) invariance. 
We neglect self-energy effects and frequency dependences. 
The additional dependence on an auxiliary energy scale $\Omega$ follows from the inclusion of a soft frequency cutoff\cite{Husemann2009} to regularize infrared divergences. 
Then, with $\Omega$ serving as flow parameter, the fRG flow equation for the two-particle coupling function reads
\begin{align} \label{eq:flow}
 \frac{d}{d\Omega}V_\Omega^{b_{1\dots 4}} &(\mathbf{k}_1,\mathbf{k}_2,\mathbf{k}_3)= \mathcal{T}_\mathrm{pp}^{b_{1\dots 4}} (\mathbf{k}_1,\mathbf{k}_2,\mathbf{k}_3) +  \\
&  + \mathcal{T}^{\, b_{1\dots 4}}_\mathrm{ph,d} (\mathbf{k}_1,\mathbf{k}_2,\mathbf{k}_3) +  \mathcal{T}^{\, b_{1\dots 4}}_\mathrm{ph,cr} (\mathbf{k}_1,\mathbf{k}_2,\mathbf{k}_3)\notag\,.
\end{align}
It involves contributions from the particle-particle ($\mathcal{T}_\mathrm{pp}$) loop, and from direct ($\mathcal{T}^\mathrm{d}_\mathrm{ph}$) as well as the crossed ($\mathcal{T}^\mathrm{cr}_\mathrm{ph}$) particle-hole loops, see Fig.~\ref{fig:diag}. 
The initial condition for the flow is given by the microscopic bare coupling $V_{\Omega_0}$, provided that the starting scale $\Omega_0$ is several orders of magnitude bigger than the bandwidth. 
Many-body instabilities towards ordered states become manifest as divergences of specific coupling components in the flow to lower energies. 
The nature of the symmetry-broken ground state is encoded by the diverging components, and the scale of divergence provides an upper estimate for the critical scale $\Omega_C$, which, in turn, can be used as an order of magnitude estimate for a gap or an ordering temperature.

The numerical implementation of the flow equation above is dealt with via the truncated-unity fRG scheme\cite{sanchez2016,Lichtenstein2016}, which follows on the methodological improvements of Refs. \onlinecite{Husemann2009,Wang2012}, and allows for the very high resolution in momentum space necessary to describe the long-ranged bare Coulomb interaction. Briefly put, there are three major manipulations carried out in Eq.~\eqref{eq:flow} leading to the TUfRG flow equations: 
\begin{enumerate}
\item The two-particle coupling is split into a bare part $V_{\Omega_0}$ and three single-channel coupling functions $\Phi^{\mathrm{P}},\Phi^{\mathrm{D}}$ and $\Phi^{\mathrm{C}}$ whose scale derivatives correspond to $\mathcal{T}_\mathrm{pp},\mathcal{T}^\mathrm{d}_\mathrm{ph}$ and $\mathcal{T}^\mathrm{cr}_\mathrm{ph}$ loops, respectively. The three original dependences of the coupling function on external momenta are rearranged in each channel so that they depend explicitly on the transfer momenta $\mathbf l$ involved in their corresponding loop diagrams. The effective coupling function may develop strong dependences on either of these transfer momenta, while having softer dependences on the remaining non-transfer momenta $\mathbf k, \mathbf{k}^\prime$.
\item The next modification is to expand the weak dependences onto a form-factor basis of lattice harmonics $\{f_n\}$. That brings each single channel coupling $\Phi^{\mathrm{B},b_{1\dots 4}}_{\mathbf l, \mathbf k, \mathbf{k}^\prime}$ to a so-called exchange propagator $B^{b_{1\dots 4}}_{m,n}(\mathbf l)$, where $B = \{P,D,C\}$. Since the weak momentum dependences can generally be captured with a small number of form-factors, in practice one is reducing a three momentum dependent function into three functions of a single momentum dependence. 
\item The last step is to insert partitions of unity in the form-factor basis set at the internal lines of the loops in Eq.~\eqref{eq:flow}, which allows to separate the fermionic Green's functions and two-particle couplings in the loop integrals. 
\end{enumerate}
Following this procedure, one arrives at the TU-fRG flow equations for the exchange propagators shown in App.~\ref{sec:tufrg}. During the flow, encountering divergences in the $P$, $D$ or $C$ channel, reflects a pairing, charge or magnetic instability, respectively. The transfer momentum $\mathbf l$ and form factor indices $m,n$ at which divergences occur reveal the ordering vector and the symmetry of the order parameter, respectively. We refer to Ref.~\onlinecite{Lichtenstein2016} for a thorough derivation of the scheme including details on computational performance and parallel scalability, or App.~\ref{sec:tufrg} for a quick overview, together with a discussion of the scheme's limitations in dealing with long-ranged interactions.
In the following, we focus on the application of the scheme to the problem at hand.

\begin{figure}[t!]
\includegraphics[width=1.0\columnwidth]{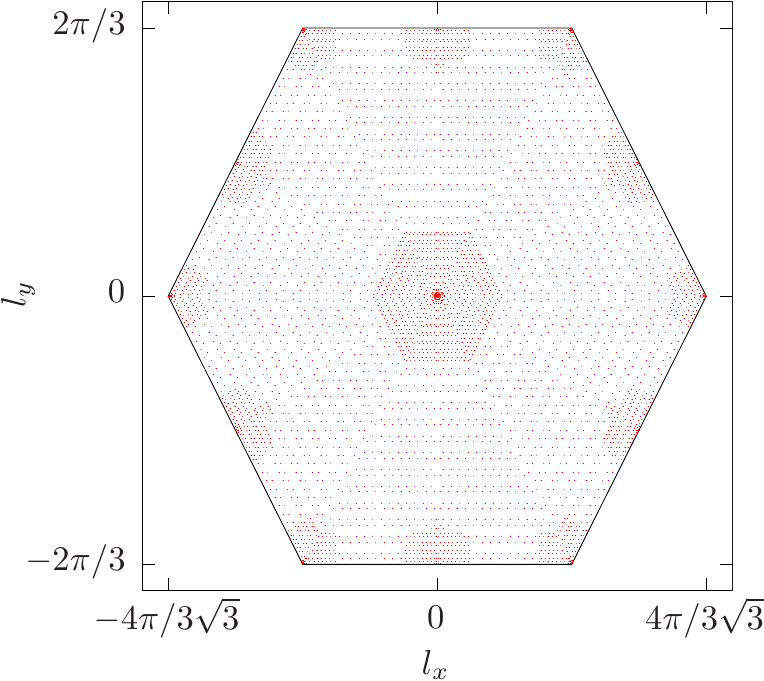}
\caption{Brillouin zone (BZ) of the honeycomb lattice with typical wave-vector discretization as implemented in the TU-fRG approach. We choose a resolution of the wave-vector discretization with a higher density around high-symmetry points, i.e. close to the center of the BZ, $\Gamma$, the corners of the BZ, $K$ and $K^\prime$, and the three M-points ($M_1,M_2,M_3$).}
\label{fig:mesh}
\end{figure}

The Brillouin zone mesh, representing the discretization of the transfer momentum, we normally use is that of Fig.~\ref{fig:mesh}, with 6097 points and a very high density around the $\Gamma$ point. They are constructed in a recursive way, starting from the irreducible $\Gamma MK$ triangle in the BZ and dividing it up into 4 similar triangles in each recursion. The minimal number of recursions is 5, the density around the M and K points corresponds to 7 and 8 recursions respectively, and over 40 recursive steps are done around the $\Gamma$ point. 

The usual number of form-factor shells considered is 3 and 4, going up to the fifth one for convergence tests. The flow equations are solved numerically with a third order Adams-Basforth multistep ODE solver \cite{Ahnert2011}, and higher order for convergence tests.

\section{Results}\label{sec:results}

We study the interaction-induced quantum-many-body instabilities of spin-1/2 fermions on graphene's honeycomb lattice in three steps: 

\begin{enumerate}
\renewcommand{\theenumi}{\Alph{enumi}}

\item Starting from a well-tested setup for the pure $t-U$ model without further interactions, or strain, we investigate for the first time the effect of a finite second-nearest neighbor hopping amplitude on the value for $U_C/t$, i.e. the critical Hubbard-interaction that induces the ordering transition to the AF-SDW. 

\item Next, we go back to the pure $t-U$ model and systematically consider extended interactions and their effect on the system stability. The short-range regime is studied starting from the local model via a stepwise inclusion of interaction terms up to a few neighbors in order to motivate the lower bound for the study of long-range interactions.

\item Finally, we connect to the lattice QMC results that have studied the ground-state of the model using different sets of \emph{ab-initio} interaction parameters and an effective long-range tail $\sim 1/(\epsilon\, r)$, cf. Ulybyshev \emph{et al.} and Tang \emph{et al.}. For that matter, we include the effect of strain to profiles given by cRPA interaction parameters and the Ohno interpolation formula. Moreover, we study the effect of a second-nearest neighbor hopping on the strained long-range interacting system.

\end{enumerate}

\subsection{Impact of second-nearest neighbor hopping -- purely local interaction}\label{sec:Ut2}

We start with the simple $t-U$ model and add a second-nearest neighbor hopping amplitude $t^\prime$ in order to better model the full band structure of graphene, cf. Eq.~\eqref{eq:sp}. 
As the second-nearest neighbor hopping is known only approximately, we sweep through a range of values for $t^\prime$ that are expected to be relevant for graphene, explicitly $|t^\prime| \in [0,0.2\,t]$. Simultaneously, for $t^\prime\neq 0$, we adjust the chemical potential such that the Fermi level lies at the Dirac point, again.

For $t^\prime=0$, employing the TU-fRG approach, we obtain a critical Hubbard interaction of $U_C\approx 2.7\, t$.
We note that this value is smaller than the numerical value of $U_{C,\mathrm{QMC}}\approx 3.8\,t$, see also the discussion in Ref.~\onlinecite{sanchez2016}.
With the instability already appearing at smaller $U_C/t$, the TU-fRG seems to overestimate the effect of fermionic fluctuations.
Another effect that tentatively increases the value of $U_C/t$ is the logarithmic renormalization of the Fermi velocity\cite{2011NatPh...7..701E}. This effect is not included within our truncation scheme since we do not take into account the flow of the self-energy. Therefore, we note that we do not expect our results to be quantitatively precise, but nevertheless, we can give estimates for parameter trends.

\begin{figure}[t!]
\includegraphics[width=1.0\columnwidth]{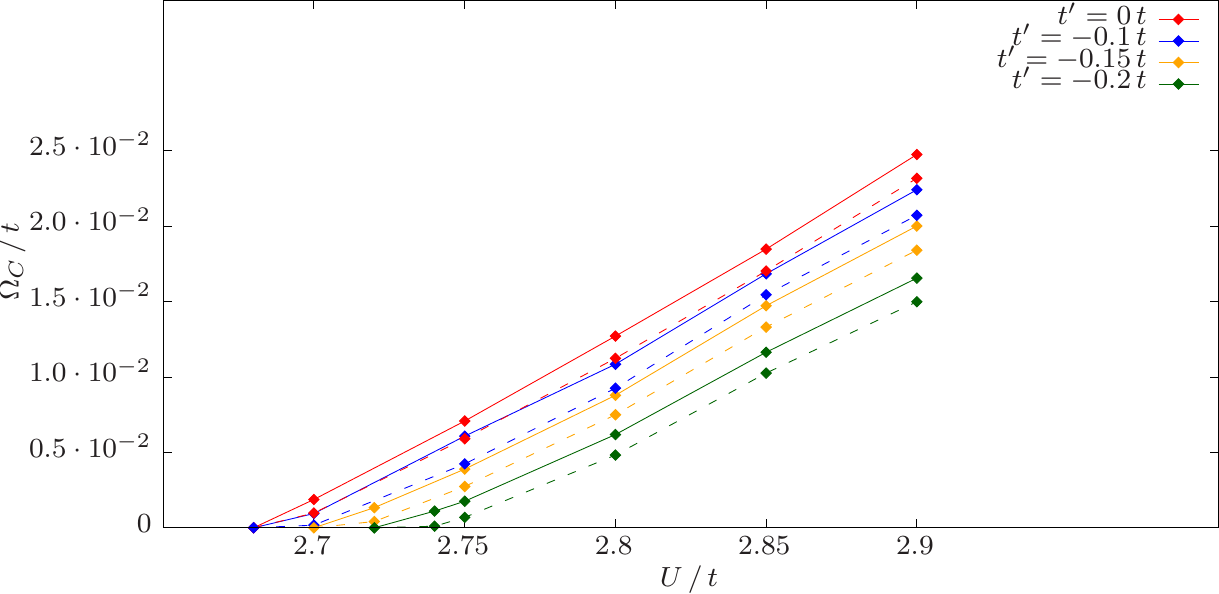}
\caption{Effect of $t^\prime$ on the critical scale $\Omega_C$ for a model with pure onsite interaction and truncation at the second shell of form-factors (solid) and at the third shell of form factors (dashed). Results including the fourth form-factor shell do not deviate from those including only till the third, indicating good convergence.}
\label{fig:uctprime}
\end{figure}

We go on to study the impact of $t^\prime$ on the critical scales $\Omega_C$ of the Hubbard-model which we interpret as an estimate for the typical gap size of the system, see Fig.~\ref{fig:uctprime}. 
We observe, that a finite $t^\prime$ does not significantly change the value of the critical onsite interaction.
This can be rationalized as close to the critical interaction, the instability will only appear for small scales and therefore is governed by the dispersion close to the Dirac points. This dispersion is not changed by the presence of $t^\prime$ except for the shift in the Fermi level, which we have absorbed by adjusting the chemical potential.
On the other hand, a finite $t^\prime$ changes the critical scales above $U_C$ considerably, cf. Fig.~\ref{fig:uctprime}. We therefore predict that a finite second-nearest neighbor hopping $t^\prime$ has a sizeable impact on the expected size of the many-body mass gaps and transition temperatures. 
For example, for $U/t =2.85$, the critical scale $\Omega_C/t$ is reduced by about 40\% upon inclusion of a second-nearest neighbor hopping $t^\prime=-0.2t$, suggesting smaller gaps than the one that would be predicted in a simple tight-binding model with nearest-neighbor hopping only.

This suppression of the critical scale is an effect beyond mean-field or single-channel ladder summations. For comparison, switching off the particle-particle channel, i.e. resorting to an effective single-channel resummation, and setting $U = 2.85 \,t$, a second-nearest neighbor hopping $t^\prime = -0.2 \, t$ leads to only a 4\% critical scale reduction respect to $t^\prime = 0$. 
The result is the same for all truncations up to the 4$^{\text{th}}$ form-factor shell, since without particle-particle channel, there is no significant inter-channel feedback and there is a fast convergence with respect to the number of form-factor shells. That smaller suppression of the critical scale is due to the breaking of particle-hole symmetry. At $t^\prime =0$, the whole Brillouin zone is perfectly nested with respect to interband scatterings with zero momentum transfer in the magnetic channel. A finite $t^\prime$ respects the approximate particle-hole nesting around the Fermi level, and therefore its influence on the critical scale is mild in terms of the particle-hole channel alone. On the other hand, it flattens the lower energy band, leading to higher particle-particle correlations, which are known to inhibit magnetism \cite{Kanamori}. We therefore conclude that the strong suppression of about 40\% seen in Fig.~\ref{fig:uctprime} is a consequence of the interplay between different channels.
%

\subsection{cRPA parameters without strain}\label{sec:short}

Starting again from the simple $t-U$ model, we next add non-local repulsive terms in a stepwise fashion, using cRPA interaction parameters \cite{PhysRevLett.106.236805} available till the fourth nearest neighbor, and extrapolating them up to the twentieth neighbor. 
A nearest-neighbor repulsive coupling $V_1$ triggers a CDW where occupancy alternates between sublattices, and a $V_2$ coupling induces a modulated charge density wave with tripled unit cell. The interplay among these coupling terms caused some controversy regarding possible exotic ground states, hinting towards spin liquid and topologically non-trivial phases\cite{raghu2008}, where most studies focused on the case of spinless fermions. 
However, in more recent studies they are falling out of favor for the more mundane charge order, both in the spinless\cite{PhysRevB.92.155137} as well as in the spin-1/2 case\cite{volpez2016,sanchez2016}.
As shown in Ref.~\onlinecite{sanchez2016}, results from our current method do not support exotic phases either, but the high momentum resolution allowed to see novel incommensurate charge ordering tendencies instead. These arise due to competition effects, with the charge ordering patterns triggered by first and second nearest neighbor interactions being incompatible and the system entering geometrical frustration.

\begin{figure}[t!]
\includegraphics[width=1.0\columnwidth]{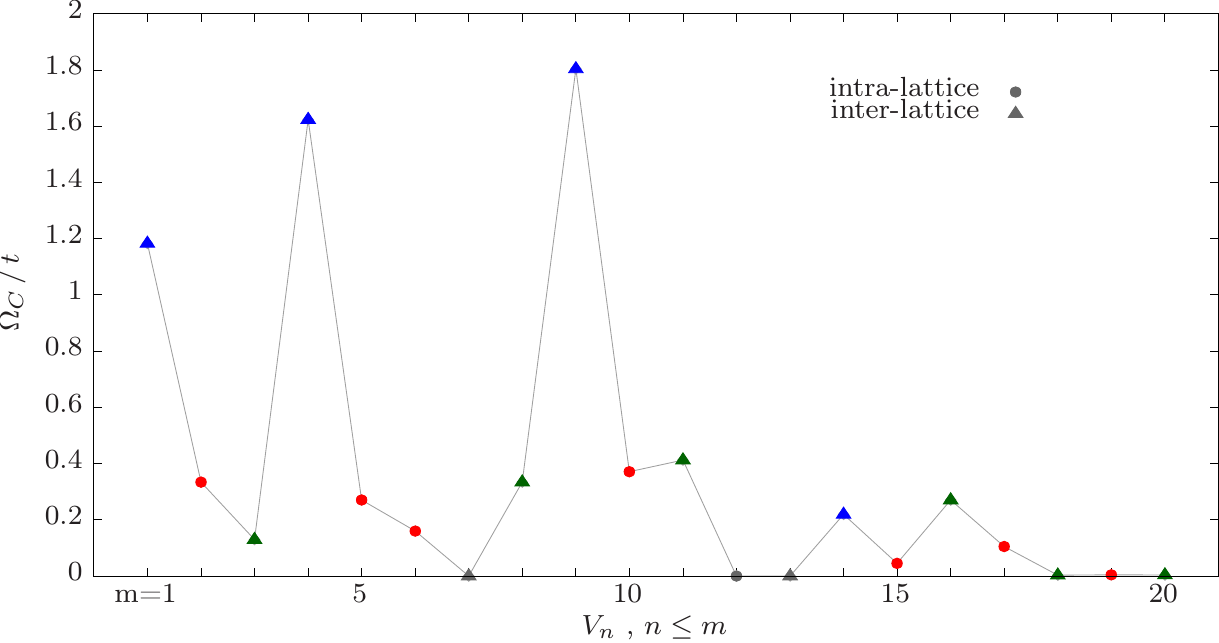}
\caption{Critical scales vs.~the number of considered nearest neighbor interactions. Inter-lattice terms are marked by triangles, whereas intra-lattice terms are represented by circles. For inter-lattice terms, a further distinction is made depending on the location of the leading ordering vector. Blue triangles correspond to charge ordering tendencies with an ordering vector close to the $\Gamma$-point, with the point $m=1$ being a standard CDW and the only commensurate case. Green triangles are ordering tendencies with ordering vectors anywhere in the BZ other than $\Gamma$. Grey points correspond to semimetallic behavior and the absence of an instability. }
\label{fig:short}
\end{figure}

Adding further agents to the competition, i.e. other non-local density-density interaction terms $V_i$ with $i>2$, reveals a rich and complex landscape of charge ordering instabilities, interspersed by points where the system remains semimetallic due to the charge ordering tendencies being balanced and suppressing magnetism. Here, we used cRPA interaction parameters as a reference.
The complex picture obtained is expected to extrapolate to other realistic parameter choices on a qualitative level.
Results are shown in Fig.~\ref{fig:short}, which can be viewed as a path in a 20-dimensional parameter space, starting from just onsite and first nearest neighbor cRPA parameters, and each step being taken in a new coupling direction. When considered alone, pure $n^{\text{th}}$-nearest-neighbor interactions result in different tendencies depending on whether the interaction is inter-lattice or intra-lattice. Inter-lattice repulsive terms are all equivalently minimized by the standard CDW, together with more complex patterns for interactions other than $V_1$, which are usually sub-leading due to degeneracy. Intra-lattice repulsive terms each support differently modulated charge density waves, with tripled, 9x, and 12x extended unit cells for pure $V_2$, $V_5$, and $V_6$ terms respectively, to name some examples. The rich interplay that arises when considered together shows ordering vectors and critical scales going back and forth, from situations which are very unstable towards incommensurate charge order with a modulation close to that of the CDW, down to situations where the semimetal remains stable. The high critical scales take place when there is a big majority of inter-lattice terms, since they all have the CDW as common tendency. The ordering vectors may lie very close to the $\Gamma$~point, but due to the presence of other tendencies they stay incommensurate. When the situation is better balanced and the scales drop, ordering vectors may appear anywhere in the BZ. For instance, as seen in Fig.~\ref{fig:short}, adding a $V_3$ coupling yields a lower $\Omega_C$ than for $m=2$, and an ordering vector close to the $K$-point. Even though it supports the CDW, $V_3$ also triggers stripe ordering patterns, manifest as sub-leading peaks in the charge propagator which are 3-fold degenerate and not dominant, but still take part in the competition. To highlight the complexity of the interplay, it must be mentioned that the CDW triggering tendencies in $V_1$ and $V_3$ do actually reinforce each other, as critical scales are higher if they are considered together rather than separate, and with all other couplings set to zero. In contrast, if $U$ and $V_2$ are not set to zero, the additional tendencies brought by $V_3$ to the interplay lead to a lower critical scale.

This analysis is meant to motivate our choice for the lower bound of the Coulomb tails considered next. We include at least interactions up to the $50^{\text{th}}$ neighbor, where charge order effects are sufficiently suppressed to have a robust semimetal. This is the case for both choices of interaction parameters, either from cRPA or from the Ohno interpolation formula. On the other hand, although the discussion of this intermediate-range physics might not be directly relevant to strained graphene, it is of importance in the context of cold atoms trapped in optical lattices, where this rich charge order landscapes may be physically realized.

Here, we add a short technical discussion of the RG flow for a stable semimetal, before going into the study of strain-induced instabilities. 
At $T=0$ and $t^\prime = \mu = 0$, Coulomb interactions stay unscreened due to the vanishing DOS at the Fermi level. 
Using the soft frequency $\Omega$-regulator of Ref.~\onlinecite{Husemann2009}, the intra-band particle-hole bubble with zero momentum transfer is suppressed by the regulator itself when $\Omega$ is large, whereas for small $\Omega$ the vanishing DOS brings it down. The inter-band particle-hole bubble does not play a qualitatively relevant role for charge screening \cite{PhysRevB.75.205418}, and thus we focus on the intra-band components in this discussion.
The particle-hole bubble's behavior in flows with the $\Omega$-regulator is shown in App.~\ref{app:unscr}.
In the TU-fRG flow equations, cf.~Eqs.~\eqref{eq:TUfRG_flow}, the bubbles involved are differentiated respect to $\Omega$. 
These exhibit a sign change at $\Omega \approx 0.63t$, where the bubble has an extremum. Thus, the Coulomb interaction experiences screening in the flow for $\Omega > 0.63t$, followed by anti-screening as $\Omega$ goes to zero, reconstructing the unscreened bare interaction one had for $\Omega \rightarrow \infty$. 
This works out well for single-channel flows with the charge channel only which is equivalent to RPA. 
However, in the full flow with all three channels, the additional contributions from inter-channel feedback may prevent the neat reconstruction of the bare interaction, which either saturates to a screened interaction, or overshoots and becomes fully unscreened for a finite $\Omega$. Whether it saturates or overshoots depends very sensitively on the choice of parameters, and the order of the ODE solver and step size. 
Therefore, this effect is most likely a numerical artefact, since we are attempting to obtain a divergent solution using explicit ODE solvers, which lack A-stability. 
It is thus unsurprising that inaccuracies in the inter-channel feedback, mainly due to form-factor basis truncation, may lead to more severe accumulated inaccuracies in the charge screening behavior. 
The latter mainly happens near critical values for a magnetic instability in the presence of long-ranged charge correlations (see grey areas in the phase diagrams of next section). In such situations we cannot flow below scales of $\Omega \sim 10^{-3}-10^{-2}t$ without encountering numerical overflows in the charge channel, due to the overestimated anti-screening. For more details about computational complications we refer to the appendix. 

\subsection{Effects of strain}

%
\begin{figure}[t!]
\includegraphics[width=1.0\columnwidth]{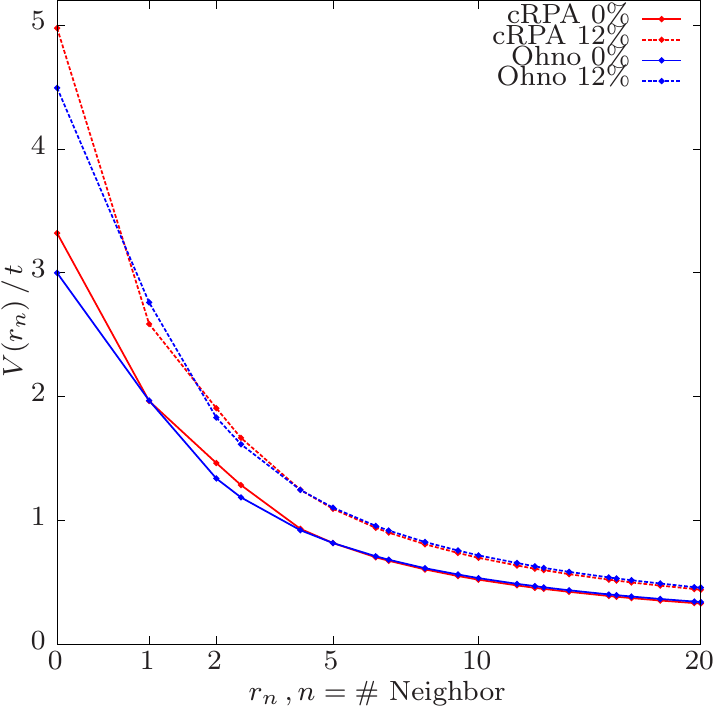}
\caption{cRPA and Ohno interaction profiles for $\lambda=1/r_{10^3}$ with 0\% and 12\% strain.}
\label{fig:V}
\end{figure}

The Brillouin zone meshes used allow to resolve interaction profiles including beyond the $10^6$-th nearest neighbor. To parameterize the interaction range, instead of including different number of neighbors as done in Sec.~\ref{sec:short}, all terms up to the  $10^4$-th nearest neighbor are considered and an artificial screening factor $e^{-\lambda r}$ is multiplied to the potential to smoothly switch off the long-range tail at the indicated number of nearest-neighbor interactions, i.e. $\lambda=1/r_n$ with $r_n$ being the distance to the furthest interaction parameter. Further ranged profiles are considered whenever critical strain values do not converge before $\lambda=1/r_{10^4}$.
Strain is accounted for as described in Sec.~\ref{sec:int_param}, with example profiles shown in Fig.~\ref{fig:V}. On the last subsection, we study the effect of including a finite second-nearest-neighbor hopping.

\subsubsection{cRPA parameters with strain}

\begin{figure}[t!]
 \includegraphics[width=\columnwidth]{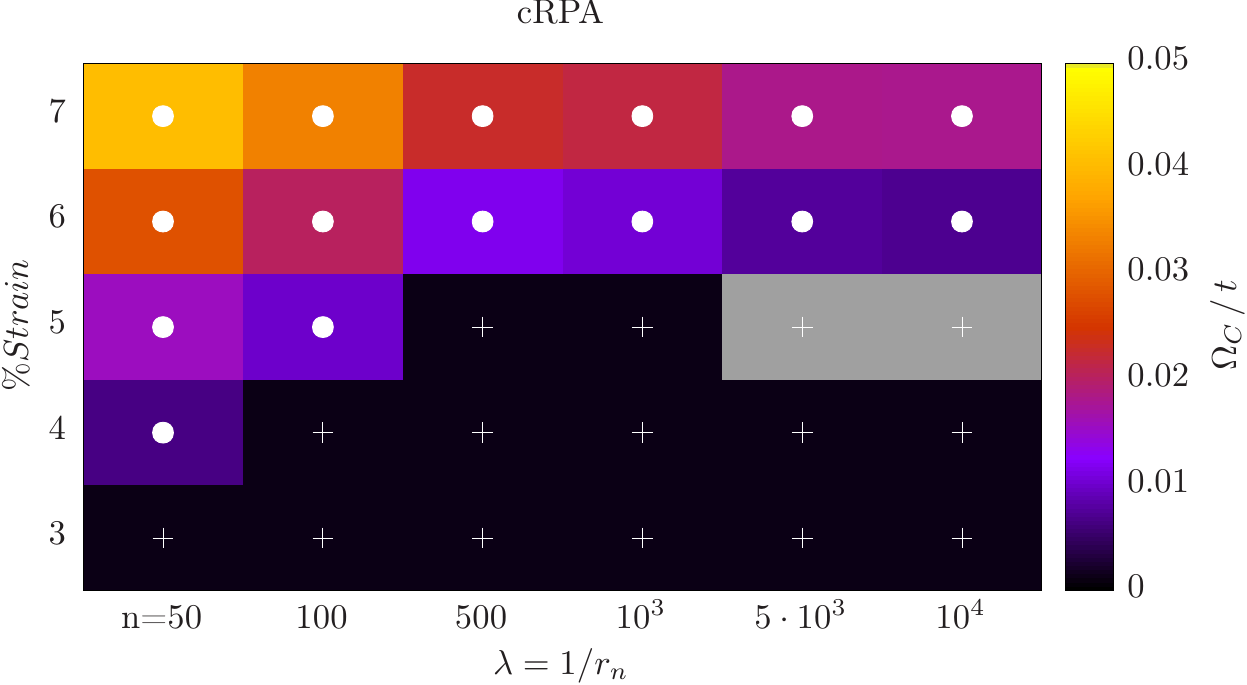}
 \caption{Effect of strain on the electronic instabilities of the model with cRPA interaction parameters. The horizontal axis denotes an artificial screening length set at the $n$-th neighbor's bond distance. The black regions marked with white crosses represent the semimetallic behavior. Filled white circles indicate an instability towards a SDW-AFM, with corresponding critical scales encoded in the background color. Grey regions are expected to stay semimetallic, but unfortunately we cannot flow down to low enough scales for those points. See text for further details.}
 \label{fig:cRPAstrain}
\end{figure}

Setting $t^\prime$ and $\mu$ to zero on our model parameters, we employ a cRPA interaction profile and study the effect of finite strain $\eta$ on the system's many-body instabilities. The concrete values used are the same as in Refs. \onlinecite{PhysRevLett.106.236805,PhysRevLett.115.186602}
We find that long-ranged cRPA interaction profiles give rise to an antiferromagnetic SDW instability for a strain larger than a critical value, see Fig.~\ref{fig:cRPAstrain}.
The critical strain necessary to induce the instability converges with respect to the inclusion of yet longer ranged Coulomb tails, staying at 6\% for profiles ranging up to the $10^5$-th neighbor and a corresponding $\lambda=1/r_{10^5}$.
Importantly, we observe that this type of interaction profiles does not give rise to other leading instabilities, but the AF-SDW, i.e. no charge ordering tendencies dominate the phase diagram.
We have checked that our results are robust with respect to denser wave-vector meshes, the inclusion of a fifth form-factor shell, or the use of a fifth order ODE solver. 
The dominance of the AF-SDW ordering tendency agrees well with findings from the QMC simulations on a qualitative level.
Based on our earlier considerations within the honeycomb-Hubbard model, cf. Sec.~\ref{sec:Ut2}, we expect that our approach overestimates the effects from fermionic fluctuations and therefore gives rise to an underestimated critical strain.
This expectation agrees with the result from the QMC calculations where for the cRPA parameters no semi-metal insulator transition could be observed for strains up to 18\%.

We note, that there is some ambiguity in the initialization procedure, relating to which channel contains the on-site Hubbard contribution: The most neutral or unbiased choice is to assign $1/3$ of it to each of the three channels, resulting in the phase diagram presented here. However, other formally equivalent ways to initialize the onsite term are expected to yield similar results, and we consider them as a consistency check. If the onsite Hubbard $U$ is fully assigned to the magnetic channel, one introduces some bias towards magnetism and obtains a critical strain of 3\% for the longer ranged profiles. In contrast, if $U$ is fully assigned to the charge channel instead, a critical strain of 10\% is obtained for long ranged profiles. A more detailed discussion of this issue can be found in App.~\ref{sec:app_b}. 
The qualitative picture that the cRPA interaction profile gives rise to an AF-SDW transition beyond a critical strain is nevertheless the same, independent of initialization.

\subsubsection{Ohno formula and strain}

 \begin{figure}[t!]
 \includegraphics[width=\columnwidth]{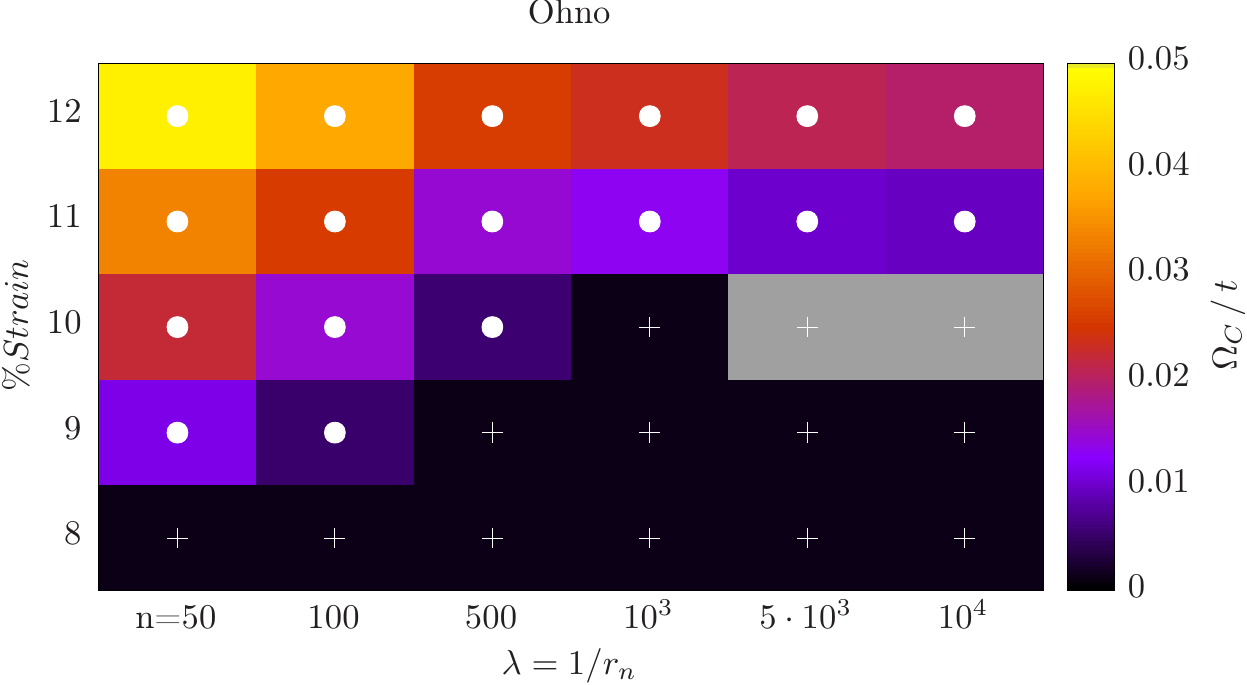}
 \caption{Effect of strain on the electronic instabilities of the model with Ohno interaction parameters. The horizontal axis denotes an artificial screening length set at the $n$-th neighbor's bond distance. The black regions marked with white crosses represent the semimetallic behavior. Filled white circles indicate an instability towards a SDW-AFM, with corresponding critical scales encoded in the background color. Grey regions are expected to stay semimetallic, but unfortunately we cannot flow down to low enough scales for those points. See text for further details.}
 \label{fig:Ohnostrain}
\end{figure}

Next, setting again $t^\prime$ and $\mu$ to zero on our model parameters, we study Ohno interaction profiles with finite strain~$\eta$ which remained elusive to the QMC calculations. 
We set the unstrained values in Eq.~\eqref{eq:ohno} to $U/t=3.0$, and choose $\epsilon$ so that $V_1/t=2.0$, then proceed analogously to the previous subsection, cf.~Fig.~\ref{fig:Ohnostrain}. The choice of a slightly smaller $U$ than in cRPA is purposely done for contrast, keeping a similarly strong non-local tail. Also note that under strain, the cRPA parameters tend faster towards a localized interaction than the Ohno parameters. 
This leads to a considerably larger critical strain for this interaction profile as compared to the strained cRPA parameters. In fact, the critical strain necessary to induce an instability converges to 11\% when including up to the $10^5$-th neighbor in the interaction.
Also, in this case no leading instability other than the AF-SDW appears.
Our results are as well robust respect to the use of denser wave-vector meshes, the inclusion of a fifth form-factor shell, or the use of a fifth order ODE solver. 
Again, there is some ambiguity in the initialization procedure. The results presented in Fig.~\ref{fig:Ohnostrain} correspond to the most neutral or unbiased choice, distributing the onsite $U$ contribution equally among the three channels. With the on-site Hubbard $U$ fully contained in the magnetic channel, we get a critical strain of 8\% for the longer ranged profiles, and if $U$ is fully assigned to the charge channel instead, the critical strain is 15\%.
As a general trend, we observe that a more strongly pronounced long-range tail in the interaction profile tends to increase the critical strain required to induce a semi-metal-insulator transition or, in other words, it stabilizes the semi-metallic behavior of the graphene model.

\begin{figure}[t!]
 \includegraphics[width=\columnwidth]{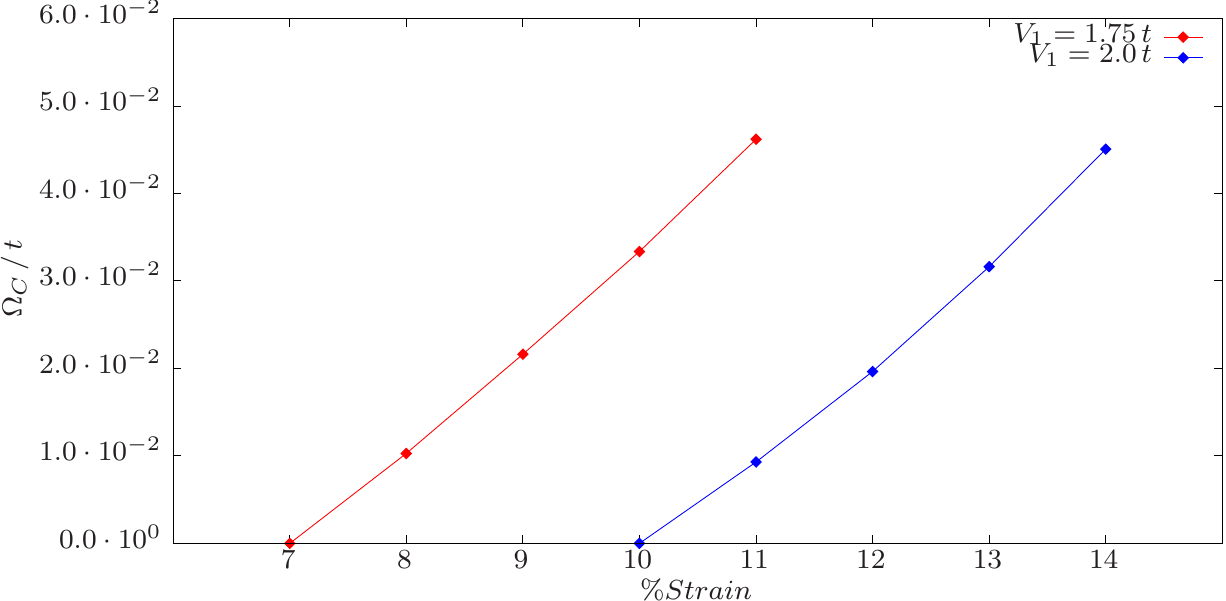}
  \caption{Critical scales vs.~strain for different interaction strengths of the Coulomb tail in the Ohno formula.}
 \label{fig:OhnoV}
\end{figure}

We also consider deviations from the model parameters used so far, with the aim of testing the qualitative robustness of the SM to AF-SDW transition indicated by our instability analysis. 
We find that slight modifications of $\epsilon$ in Eq.~\eqref{eq:ohno} result in a shift of the critical strain, but does not change the nature of the instability, i.e. the tendency towards the AF-SDW instability prevails.
In Fig.~\ref{fig:OhnoV}, we exhibit the effect of increasing $\epsilon$ to a value that yields $V_1/t=1.75$, resulting in a smaller critical strain. This is in agreement with our earlier observation since the larger value of $\epsilon$ leads to a less-pronounced long-range tail. 
Setting smaller $\epsilon$'s aggravates the aforementioned technical difficulties that arise with the unscreening of charge interactions. For instance, when choosing $\epsilon$ such that $V_1/t=2.25$, we can only say that the critical strain is shifted to about 15-18\%, but cannot give a more definite answer. 
Instead of rising the non-local terms to higher values, we can alternatively lower the on-site interaction strength. If we set $U/t=2.5$, keeping the rest unchanged, we obtain an instability to incommensurate charge order for strains above 10-15\%, and recover the antiferromagnet when strain reaches about 30\%. In these comparisons, one has to push the ratio between on-site and extended terms to unrealistic values in order to trigger instabilities other than the AF-SDW. This is due to the fact that the Coulomb tail is modified as a whole, which does not sufficiently disturb the balanced competition among charge ordering tendencies. 
However, if we disturb that balance, charge order is much more likely to appear. In the original set of parameters, with $U/t=3.0$ and $\epsilon = 1.25$, it suffices to increase $V_1/t = 2.0$ to $V_1/t = 2.25$ while keeping the rest unchanged, to make even the unstrained system unstable towards an iCDW. The quantitative impact of this deviations has not been tested for convergence.

\subsubsection{$t$-$t^\prime$-Hubbard-Coulomb model with strain}

\begin{figure}[t!]
 \includegraphics[width=\columnwidth]{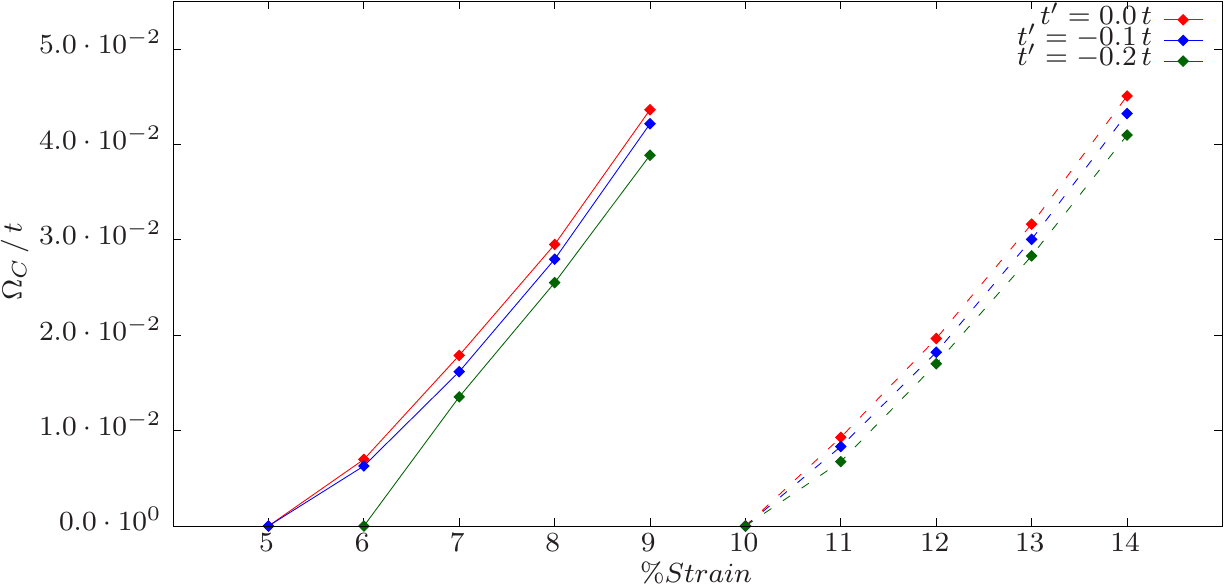}
  \caption{Critical scales vs.~strain for different values of the second-nearest-neighbor hopping with cRPA (solid) and Ohno (dashed) interaction parameters.}
 \label{fig:full}
\end{figure}

Finally, we study the full model Hamiltonian to explore a close-to-realistic model for graphene. To that end, we include a second-nearest neighbor hopping as well as the two interaction profiles from the cRPA and the Ohno method and investigate the effect of a finite amount of strain. Explicitly, we compare the critical scales for the appearance of a many-body instability for three different choices of the unstrained second-nearest neighbor hopping $t^\prime \in\{0,-0.1t,-0.2t\}$. We note that the application of strain quickly reduces the value of the second-nearest neighbor hopping $t^\prime$ following the relation in Eq.~\eqref{eq:hopstrain}, while increasing the interaction strength relative to $t$.
Therefore, we expect a smaller impact of $t^\prime$ on the critical scales as compared to the pure modification of the Hubbard interaction as studied in Sec.~\ref{sec:Ut2}. 
The results of this study are shown in Fig.~\ref{fig:full} and confirm this expectation.
For both, the cRPA as well as the Ohno interaction profiles, the critical scales for different values of strain only weakly depend on the chosen unstrained value of the second-nearest neighbor hopping $t^\prime$.

Finally, we comment on a suggestion for an effective honeycomb Hubbard model derived from the honeycomb Hubbard-Coulomb model as put forward by Sch\"uler {\it et al} in Ref.~\onlinecite{PhysRevLett.111.036601}. Noting that non-local charge interactions stabilize the Dirac semimetal against magnetic ordering, and provided the absence of other instabilities, they proposed a pure on-site Hubbard model with downscaled local interaction $U^\ast$ as a reasonable approximation. More specifically, one would then have $U^\ast = U - \bar{V}$, where $\bar{V}$ is a weighted average of non-local terms which they further approximate by $\bar{V} \approx V_1$. Concrete values for the interaction profiles considered in this work can be found in Table \ref{tab:wehling}, where a common trend of $U^\ast_C \approx 1.75 \, t$ can be inferred, which is to be compared to the $U_C \approx 2.7 \,t$ of the original local Hubbard model. Nonetheless, quantitative differences aside, our results support the validity of an effective on-site model. The crucial aspect here is the absence of leading ordering tendencies other than antiferromagnetism. 

\begin{table}[t!]
\caption{\label{tab:wehling} Effective Hubbard repulsion according to Ref.~\onlinecite{PhysRevLett.111.036601} for the different interaction profiles at different values of strain. Ohno 1 and 2 denote the two different profiles considered in Fig.~\ref{fig:OhnoV} with stronger and weaker Coulomb tails, respectively. See the text for further discussion.}
\begin{tabular*}{\linewidth}{@{\extracolsep{\fill} } c c c c c c}
\hline\hline
i.a. profile  & strain & $U/t$ & $V_1 /t$ & $U^\ast/t$ & instability \\
\hline
cRPA & 0\% & 3.3 & 2.0 & 1.3 & x \\
cRPA & 6\% & 4.05  & 2.25  & 1.8  & $\checkmark$ \\
cRPA & 12\% & 5.0 & 2.6 & 2.4 & $\checkmark$ \\
\hline
Ohno 1  & 0\% & 3.0 & 2.0 & 1.0 & x \\
Ohno 1  & 12\% & 4.5 & 2.75 & 1.75 & $\checkmark$ \\
\hline
Ohno 2  & 0\% & 3.0 & 1.75 & 1.25 & x \\
Ohno 2  & 8\% & 3.9 & 2.15  & 1.75 & $\checkmark$ \\
\hline\hline
\end{tabular*}
\end{table}
%

\section{Conclusions}\label{sec:conc}

In the following, we summarize our main results. First, we find a sizable reduction of up to 40\% for the critical scales in the antiferromagnetic transition of the honeycomb-Hubbard model upon inclusion of a finite second-nearest neighbor hopping $t^\prime$ chosen within the range of suggested {\it ab initio} values. 
We showed that this effect is beyond single-channel resummations and results from the complex interplay between different interaction channels.
This finding suggests that a finite $t^\prime$ causes a considerable reduction of expected gap sizes in the honeycomb-Hubbard model.

Furthermore, the consecutive inclusion of more and more remote non-local interaction terms up to the $20^{\text{th}}$-nearest neighbor following the unstrained cRPA interaction profile provides a sequence of different (in-)commensurate charge ordering patterns. 
The critical scales of these charge orders discontinuously jump from rather large values to zero and back, indicating a strong competition between these orders. 
Magnetic instabilities are suppressed.
When including enough interaction terms, the competition between the charge ordering patterns drives the system into a frustrated regime where no instability appears and semi-metallic behavior prevails.
We conclude that the semi-metallic behavior of graphene is not a result of the smallness of interactions but due to a strong competition and an eventual frustration of different ordering tendencies.

This frustration can be lifted by application of a biaxial strain which we have studied by employing two different types of long-ranged interaction profiles, i.e. the cRPA and the Ohno interpolation to take account for the uncertainties in the determination of interaction parameters.
We showed that for both the cRPA as well as the Ohno profiles, a critical strain exists beyond which the system develops a quantum many-body instability. The TU-fRG values for the critical strain lie between about 5\% (cRPA) and 11\% (Ohno).
Notably, the nature of the leading instability for these long-ranged interaction profiles is of AF-SDW type, i.e. charge ordering tendencies are never preferred despite their importance for intermediate-range potentials. This option could not be explored before, as the QMC calculations typically suffer from a sign-problem for interaction potentials with a strong tail. The nature of the possible instabilities turn out to be the same for both pure on-site and long-ranged interacting models, which also persists under inclusion of a finite second-nearest neighbor hopping term. Thus, this is supporting evidence for the qualitative validity of effective honeycomb $t$-$U$-Hubbard models in place of $t$-$t^\prime$-Hubbard-Coulomb models. 

Generally, the results of the TU-fRG approach presented here, overestimate the effect of fermionic fluctuations which leads to an earlier onset of ordering tendencies. We conjecture that this is in part caused by the neglect of self-energy effects which would, for example, lead to finite lifetime effects \cite{PhysRevLett.77.3589,Honerkamp2017a,Honerkamp2017b} and the renormalization of the Fermi velocity \cite{2011NatPh...7..701E}. Therefore, for more quantitative estimates, an inclusion of self-energy effects within the fRG approach would be desirable. We expect this task to be numerically demanding but feasible in the future.

\begin{acknowledgments}
The authors are grateful to T.~O.~Wehling, F.~F.~Assaad and J.~N.~B~Rodrigues for discussions and sharing their data. We also thank T.~Reckling and Q-H.~Wang for useful discussions. We acknowledge support by the DFG priority program SPP1459 on graphene and the research unit RTG1995. 

The authors also acknowledge the use of the ODEint library \cite{Ahnert2011}, the DCUHRE routine \cite{Berntsen1991} and the JUBE environment \cite{Luhrs2016}.

The authors gratefully acknowledge the computing time granted by the JARA-HPC Vergabegremium on the supercomputer JURECA at Forschungszentrum J\"ulich.
\end{acknowledgments}

\onecolumngrid

\appendix

\section{Truncated-unity functional Renormalization Group scheme}\label{sec:tufrg}
In the presence of U(1), SU(2) and translational invariance, the flow equation for the two particle coupling function in the level-two truncation of the hierarchy for 1PI vertices reads
\begin{equation*}
\partial_\Omega V^{b_{1\dots 4}} (k_1,k_2,k_3) = \mathcal{T}_\mathrm{pp}^{b_{1\dots 4}} (k_1,k_2,k_3) + \mathcal{T}^{\mathrm{cr}, \, b_{1\dots 4}}_\mathrm{ph}  (k_1,k_2,k_3) + \mathcal{T}^{\mathrm{d}, \, b_{1\dots 4}}_\mathrm{ph} (k_1,k_2,k_3) 
\end{equation*}
where $k_i = (\omega_i,\mathbf{k}_i)$, dependences on the regularization scale $\Omega$ are implicitly understood. The three contributions in the right hand side read
\begin{align} \label{eq:diag}
 \mathcal{T}_\mathrm{pp}^{b_{1\dots 4}} = - \int \! d p & \, \left[ \partial_\Omega \,G(p,b) \, G(k_1+k_2-p,b') \right]  V^{b_1 b_2 bb'} (k_1,k_2,p) \,V^{bb'b_3 b_4}(p,k_1+k_2-p,k_3)\,, \notag \\[1.0ex] 
 \mathcal{T}^{\mathrm{cr}, \, b_{1\dots 4}}_\mathrm{ph} = - \int \! d p & \, \left[ \partial_\Omega \,G(p,b) \, G(p+k_2-k_3,b') \right]  V^{b_1 b'bb_4} (k_1,p+k_2-k_3,p) \, V^{bb_3b_2 b'}(p,k_2,k_3)\,,  \\[1.0ex] 
 \mathcal{T}^{\mathrm{d}, \, b_{1\dots 4}}_\mathrm{ph} = - \int \! d p & \,\, \left[ \partial_\Omega \,G(p,b) \, G(p+k_3-k_1,b') \right] \left[ -2 V^{b_1 bb_3 b'} (k_1,p,k_3) \, V^{b'b_2 bb_4}(p+k_1-k_3,k_2,p) \right. \notag \\
&  +  V^{b_1 b' b b_3} (k_1,p,p+k_1-k_3,p) \, V^{bb_2 b' b_4}(p+k_1-k_3,k_2,p) \notag  \\  
& \left. +  V^{b_1 b' bb_4} (k_1,p,k_3) \, V^{b b_3 b_2 b'}(k_2,p+k_1-k_3,p) \right]\,.  \notag
\end{align}
where $\int d p$ is shorthand notation for $\int \frac{d \mathbf{p}}{A_{B\!Z}}\frac{1}{\beta}\sum_{\omega} \sum_{b b'}$, and $G$'s are regularized bare Green's functions including the soft frequency cutoff introduced in Ref.~\onlinecite{Husemann2009}. For a frequency independent interaction and setting external frequencies to zero, the internal frequency sum in Eqs. (\ref{eq:diag}) can be done analytically. For momentum dependences, we follow the steps outlined in Sec. \ref{sec:method}, which lead to the TUfRG scheme \cite{Lichtenstein2016,sanchez2016}. The coupling function $V$ is rewritten into a bare part $V^{\Omega_0}$ and three single-channel coupling functions
\begin{equation} \label{eq:channel_decomp}
 V^{\{b_i\}} \left(\mathbf{k}_1,\mathbf{k}_2,\mathbf{k}_3 \right) = V^{\Omega_0,\, \{b_i\}}_{\mathbf{k}_1,\mathbf{k}_2,\mathbf{k}_3}  - \Phi^{\mathrm{SC},\,\{b_i\}}_{\mathbf{k}_1+\mathbf{k}_2,\frac{\mathbf{k}_1-\mathbf{k}_2}{2},\frac{\mathbf{k}_3-\mathbf{k}_4}{2}}  + \Phi^{\mathrm{C},\, \{b_i\}}_{\mathbf{k}_3-\mathbf{k}_2,\frac{\mathbf{k}_1+\mathbf{k}_4}{2},\frac{\mathbf{k}_2+\mathbf{k}_3}{2}} + \Phi^{\mathrm{D}, \{b_i\}}_{\mathbf{k}_1-\mathbf{k}_3,\frac{\mathbf{k}_1+\mathbf{k}_3}{2},\frac{\mathbf{k}_2+\mathbf{k}_4}{2}}\,, \, 
\end{equation}
which are generated during the flow in the following way
\begin{align} \label{eq:flow_phis}
\partial_\Omega \Phi^{\mathrm{SC},\,\{b_i\}}_{\mathbf{k}_1+\mathbf{k}_2,\frac{\mathbf{k}_1-\mathbf{k}_2}{2},\frac{\mathbf{k}_3-\mathbf{k}_4}{2}}& =  - \mathcal{T}_\mathrm{pp}^{\{b_i\}}  \left(\mathbf{k}_1,\mathbf{k}_2,\mathbf{k}_3 \right)\,, \notag\\
\partial_\Omega \Phi^{\mathrm{C},\,\{b_i\}}_{\mathbf{k}_3-\mathbf{k}_2,\frac{\mathbf{k}_1+\mathbf{k}_4}{2},\frac{\mathbf{k}_2+\mathbf{k}_3}{2}} &=  \mathcal{T}_\mathrm{ph}^{\mathrm{cr}, \, \{b_i\}}  \left(\mathbf{k}_1,\mathbf{k}_2,\mathbf{k}_3 \right)\,,  \\
\partial_\Omega \Phi^{\mathrm{D},\,\{b_i\}}_{\mathbf{k}_1-\mathbf{k}_3,\frac{\mathbf{k}_1+\mathbf{k}_3}{2},\frac{\mathbf{k}_2+\mathbf{k}_4}{2}} & =   \mathcal{T}_\mathrm{ph}^{\mathrm{d}, \, \{b_i\}}  \left(\mathbf{k}_1,\mathbf{k}_2,\mathbf{k}_3 \right)\,. \notag
\end{align}
Each $\Phi$ has a strong dependence on the transfer momenta involved in its corresponding loop diagram (first argument), which will be kept explicitly, and two other dependences which will be expanded onto a basis of lattice harmonics
\begin{align}\label{eq:expansion}
 \Phi^{\mathrm{SC},\,\{b_i\}}_{\mathbf{l},\mathbf{k},\mathbf{k'}} & = \sum_{m,n} f_m(\mathbf{k}) \, f^*_n(\mathbf{k'}) \, P^{\{b_i\}}_{m,n} (\mathbf{l})\,, \\
 \Phi^{\mathrm{C},\,\{b_i\}}_{\mathbf{l},\mathbf{k},\mathbf{k'}} & = \sum_{m,n} f_m(\mathbf{k}) \, f^*_n(\mathbf{k'}) \, C^{\{b_i\}}_{m,n} (\mathbf{l})\,, \\
 \Phi^{\mathrm{D},\,\{b_i\}}_{\mathbf{l},\mathbf{k},\mathbf{k'}} & = \sum_{m,n} f_m(\mathbf{k}) \, f^*_n(\mathbf{k'}) \, D^{\{b_i\}}_{m,n} (\mathbf{l})\,.
\end{align}
After the insertion of partitions of unity in the form factor basis in-between $V$s and $G$s in Eqs. (\ref{eq:diag}), the flow equations for $\Phi$s in (\ref{eq:flow_phis}) can be rearranged into flow equations for the so-called exchange propagators $P$,$C$,$D$
\begin{align}\label{eq:TUfRG_flow}
 \dot P_{m,n}^{\{b_i\}}(\mathbf{l}) = \sum_{m',n'} \sum_{b,b'} & V^{\mathrm{P},\,b_{1}b_{2}bb'}_{m,m'}\!(\mathbf{l})\,\dot \chi^{\mathrm{pp},\,bb'}_{m',n'}\!(\mathbf{l})\,V^{\mathrm{P},\,bb'b_{3}b_{4}}_{n',n}(\mathbf{l})\,, \notag \\[1.0ex]
 \dot C_{m,n}^{\{b_i\}}(\mathbf{l}) = \sum_{m',n'} \sum_{b,b'} & -V^{\mathrm{C},\,b_{1}b'bb_{4}}_{m,m'}(\mathbf{l})\,\dot \chi^{\mathrm{ph},\,bb'}_{m',n'}\!(\mathbf{l})\,V^{\mathrm{C},\,bb_{3}b_{2}b'}_{n',n}\!(\mathbf{l})\,,  
 \end{align}
 
 \begin{align}
  \dot D_{m,n}^{\{b_i\}}(\mathbf{l}) = \sum_{m',n'} \sum_{b,b'} & \Big( 2V^{\mathrm{D},\,b_{1}bb_{3}b'}_{m,m'}\!(\mathbf{l})\,\dot \chi^{\mathrm{ph},\,bb'}_{m',n'}\!(\mathbf{l})\,V^{\mathrm{D},\,b'b_{2}bb_{4}}_{n',n}(\mathbf{l}) -V^{\mathrm{C},\,b_{1}b'bb_{3}}_{m,m'}(\mathbf{l})\,\dot \chi^{\mathrm{ph},\,bb'}_{m',n'}\!(\mathbf{l})\,V^{\mathrm{D},\,bb_{2}b'b_{4}}_{n',n}(\mathbf{l}) \notag \\[1.0ex] 
& -V^{\mathrm{D},\,b_{1}bb_{3}b'}_{m,m'}\!(\mathbf{l})\,\dot \chi^{\mathrm{ph},\,bb'}_{m',n'}\!(\mathbf{l})\,V^{\mathrm{C},\,b_{2}b'bb_{4}}_{n',n}(\mathbf{l})\Big)\,, \notag
\end{align}
with 
\begin{align} \label{eq:chis}
 \chi^{\mathrm{pp},\,bb'}_{m,n} (\mathbf{l}) &= \int \! dp \, G \left(\omega_p\,, \frac{\mathbf{l}}{2} + \mathbf{p},b \right) \, G \left(-\omega_p\,, \frac{\mathbf{l}}{2} - \mathbf{p},b' \right) \, f^*_m (\mathbf{p}) \, f_n (\mathbf{p}) \,, \notag \\[0.5ex]
 \chi^{\mathrm{ph},\,bb'}_{m,n} (\mathbf{l}) &= \int \! dp \, G \left(\omega_p\,, \mathbf{p} + \frac{\mathbf{l}}{2},b \right) \, G \left(\omega_p\,, \mathbf{p} - \frac{\mathbf{l}}{2},b' \right) \, f^*_m (\mathbf{p}) \, f_n (\mathbf{p}) \,, 
\end{align}
and channel-projected coupling functions
\begin{align} \label{eq:projs}
  V^{\mathrm{P},\,\{b_i\}}_{m,n}\left(\mathbf{l} \right) = \hat{P} \left[ V^{\{b_i\}} \right]_{m,n} (\mathbf{l})\,,\quad
  V^{\mathrm{C},\,\{b_i\}}_{m,n}\left(\mathbf{l} \right) = \hat{C} \left[ V^{\{b_i\}} \right]_{m,n} (\mathbf{l})\,, \quad
  V^{\mathrm{D},\,\{b_i\}}_{m,n}\left(\mathbf{l} \right) & = \hat{D} \left[ V^{\{b_i\}} \right]_{m,n} (\mathbf{l})\,,
\end{align}
where projection operators $\hat{P}$,$\hat{C}$,$\hat{D}$ act as an inverse to the expansions (\ref{eq:expansion}), and derivatives respect to $\Omega$ are written in dot notation. 

Starting the flow at a high enough $\Omega$, the values for the channel-projected $V$s in (\ref{eq:projs}) are the corresponding projections of the bare coupling $V^{\Omega_0}$, which take the form
\begin{align}\label{eq:init_proj}
\hat{P}\left[V^{\Omega_0,\,\{b_i\}}\right]_{m,n}(\mathbf{l}) & = \sum_{\{o_i\}}\sum_{\mathbf{R}_n^{\{o_i\}}} \tilde{V}^{\{o_i\}} \left(\mathbf{R}_n^{\{o_i\}}\right) \int \!d \mathbf{k} \,f^*_m(\mathbf{k}) \,e^{-i\mathbf{k}\cdot\mathbf{R}_n^{\{o_i\}}} \hat{T}^{b_1,o_1}_{\frac{\mathbf{l}}{2}+\mathbf{k}}\,\hat{T}^{b_2,o_2}_{\frac{\mathbf{l}}{2}-\mathbf{k}} \notag \\ 
& \times \int \!d \mathbf{k}'\, f_n(\mathbf{k}') \, e^{i\mathbf{k}'\cdot\mathbf{R}_n^{\{o_i\}}} \left(\hat{T}^{b_3,o_3}_{\frac{\mathbf{l}}{2}+\mathbf{k}'}\right)^*  \left(\hat{T}^{b_4,o_4}_{\frac{\mathbf{l}}{2}-\mathbf{k}'}\right)^* \,, \notag \\[1.5ex] 
\hat{C}\left[V^{\Omega_0,\,\{b_i\}}\right]_{m,n}(\mathbf{l}) & = \sum_{\{o_i\}}\sum_{\mathbf{R}_n^{\{o_i\}}} \tilde{V}^{\{o_i\}} \left(\mathbf{R}_n^{\{o_i\}}\right) \int \!d \mathbf{k} \,f^*_m(\mathbf{k}) \,e^{-i\mathbf{k}\cdot\mathbf{R}_n^{\{o_i\}}} \hat{T}^{b_1,o_1}_{\mathbf{k}+\frac{\mathbf{l}}{2}}\,\left(\hat{T}^{b_4,o_4}_{\mathbf{k}-\frac{\mathbf{l}}{2}}\right)^*  \\ 
& \times \int \!d \mathbf{k}' \,f_n(\mathbf{k}')\, e^{i\mathbf{k}'\cdot\mathbf{R}_n^{\{o_i\}}} \hat{T}^{b_2,o_2}_{\mathbf{k}'-\frac{\mathbf{l}}{2}}  \left(\hat{T}^{b_3,o_3}_{\mathbf{k}'+\frac{\mathbf{l}}{2}}\right)^* \,, \notag \\[1.5ex] 
\hat{D}\left[V^{\Omega_0,\,\{b_i\}}\right]_{m,n}(\mathbf{l}) & = \sum_{\{o_i\}}\sum_{\mathbf{R}_n^{\{o_i\}}} \tilde{V}^{\{o_i\}} \left(\mathbf{R}_n^{\{o_i\}}\right) e^{-i\mathbf{l}\cdot\mathbf{R}_n^{\{o_i\}}} \!\int \!d \mathbf{k}\, f^*_m(\mathbf{k}) \, \hat{T}^{b_1,o_1}_{\mathbf{k}+\frac{\mathbf{l}}{2}} \left(\hat{T}^{b_3,o_3}_{\mathbf{k}-\frac{\mathbf{l}}{2}}\right)^*\, \notag \\ 
& \times \int \! d \mathbf{k}'\, f_n(\mathbf{k}') \, \hat{T}^{b_2,o_2}_{\mathbf{k}'-\frac{\mathbf{l}}{2}}   \left(\hat{T}^{b_4,o_4}_{\mathbf{k}'+\frac{\mathbf{l}}{2}}\right)^* \,, \notag
\end{align}
\twocolumngrid
\noindent where $\tilde{V}^{\{o_i\}} (\mathbf{R}_n^{\{o_i\}})$ are the bare coupling strengths at $n$-th nearest neighbor bond vector $\mathbf{R}_n^{\{o_i\}}$ connecting the orbitals $\{o_i\}$, and $\hat{T}^{b_i,o_i}_{\mathbf{k}_i}$ are the transformation elements between orbital and band degrees of freedom, chosen as
\begin{align}\label{eq:T_k1}
 \hat{\mathbf{T}}_{\mathbf{k}} & = \frac{1}{\sqrt{2}}
\begin{pmatrix}
\frac{h(\mathbf{k})}{\left|h(\mathbf{k})\right|} & -1 \\
\frac{h(\mathbf{k})}{\left|h(\mathbf{k})\right|} & 1
\end{pmatrix} \\
h(\mathbf{k}) & = \sum_{\boldsymbol{\delta}} e^{i \mathbf{k} \cdot \boldsymbol{\delta}} 
\end{align}
where $\boldsymbol{\delta}$ are the nearest neighbor bond vectors.

The exchange propagators are zero initially, and they absorb the renormalization corrections to the bare coupling during the flow. In the usual instability analyses, the flow typically begins with a weakly coupled situation and it must be stopped as soon as coupling function components grow beyond the order of magnitude of the single-particle bandwidth. This only applies till inclusions of very few nearest-neighbor interaction terms in the bare coupling. The $\frac{1}{r}$ behavior of the Coulomb potential translates to a $\frac{1}{\lvert \mathbf{l} \rvert}$ behavior for its two dimensional Fourier transform. Since the number of neighbors in our calculation is finite, the bare coupling stays finite at zero $\mathbf{l}$, but with a high enough number of neighbors included it takes values which are well over the order of magnitude of the single-particle bandwidth. The alternative is to impose the stopping condition on the difference between renormalized and bare coupling. That way, even though the flow starts with a projected bare coupling in the charge channel exhibiting a strong peak, attention is paid to whether new sharp structures are generated during the flow. That not being the case, we interpret the result as a semi-metal.

\section{Initialization procedure}
\label{sec:app_b}
For the sake of generality and unbiasedness, the initial bare interaction $V^{\Omega_0}$ was kept as a separate term in the channel decomposition of Eq.~\eqref{eq:channel_decomp}. However, in practice it may be more convenient to assign it as initial condition in the channels, especially when dealing with Coulomb interactions. For instance, the extended density-density bare interactions considered in this work are most accurately described in the $D$ channel, whereas for the constant on-site bare term it is more natural to split the contribution equally among channels. This can be understood from Eq.~\eqref{eq:init_proj} if one momentarily ignores the multi-orbital case. In a one-band situation, a Coulomb interaction projected onto the $D$ channel is fully contained in the $\mathbf{l}$ dependence of the on-site form-factor components. In contrast, the projection of a Coulomb interaction into the $P$ and $C$ channels takes the form
$\sum_{\mathbf{R}_n} \frac{1}{\lvert \mathbf{R}_n \rvert} \int \!d \mathbf{k} \,f^*_m(\mathbf{k}) \,e^{-i\mathbf{k}\cdot\mathbf{R}_n} \int \!d \mathbf{k'} \,f_n(\mathbf{k'}) \,e^{i\mathbf{k'}\cdot\mathbf{R}_n}$,
which after the $\mathbf{R}$ sum, yields non-zero contributions for all diagonal terms in the form-factor indices, with no dependence in $\mathbf{l}$. These values correspond to the interaction strength at the distance where the form-factors are defined in real space. Thus, a given interaction would need as many form-factors as lattice positions covered by its range, in order to be completely captured in each and every channel. This is not feasible for long-ranged interactions, and it suffices to have them properly captured in one channel, and truncated in the remaining two. We normally include form-factors covering at least till the $10^{\text{th}}$ nearest-neighbor, so that channels other than $D$ still get their fair share of the Coulomb interaction, though truncated. In the two-band situation considered in this work, the discussion above holds, although non-diagonal form-factor components take finite but small initial values. 

Going back to the original point, the main problem that arises when keeping $V^{\Omega_0}$ separate in the decomposition is that $V^P$ and $V^C$ often display spurious behavior. In this decomposition, $D$ collects big counter-terms to $V^{\Omega_0}$ when the effective charge interaction gets screened. In turn, the feedback from $D$ into the other two channels should also counter the respective projections of $V^{\Omega_0}$, so that all three projected $V$'s describe the same screened interaction. However, due to inaccuracies in the inter-channel projections (mainly the form-factor basis truncation) some $V^P$ and $V^C$ components remain unbalanced, vertex symmetries are not satisfied, and the flow usually ends up signaling unphysical instabilities when interactions get unscreened. Instead of separating bare couplings and their renormalized corrections, and having to rely on their accurate counterbalancing in all channels, keeping them together results in more numerically stable flows. Furthermore, one avoids computing some challenging integrals in Eq.~\eqref{eq:init_proj}, since extended interactions are all put into the $D$ channel, where the complex exponential containing $\mathbf{R}$ vectors lies outside the integrals. Initializing extended interactions in the other two channels involves integrating functions of $\mathbf{R}$, which do not have the periodicity of the reciprocal lattice in a bipartite lattice system, and lead to discontinuities when back-folded into the first BZ.

\begin{figure}[t!]
\includegraphics[width=1.0\columnwidth]{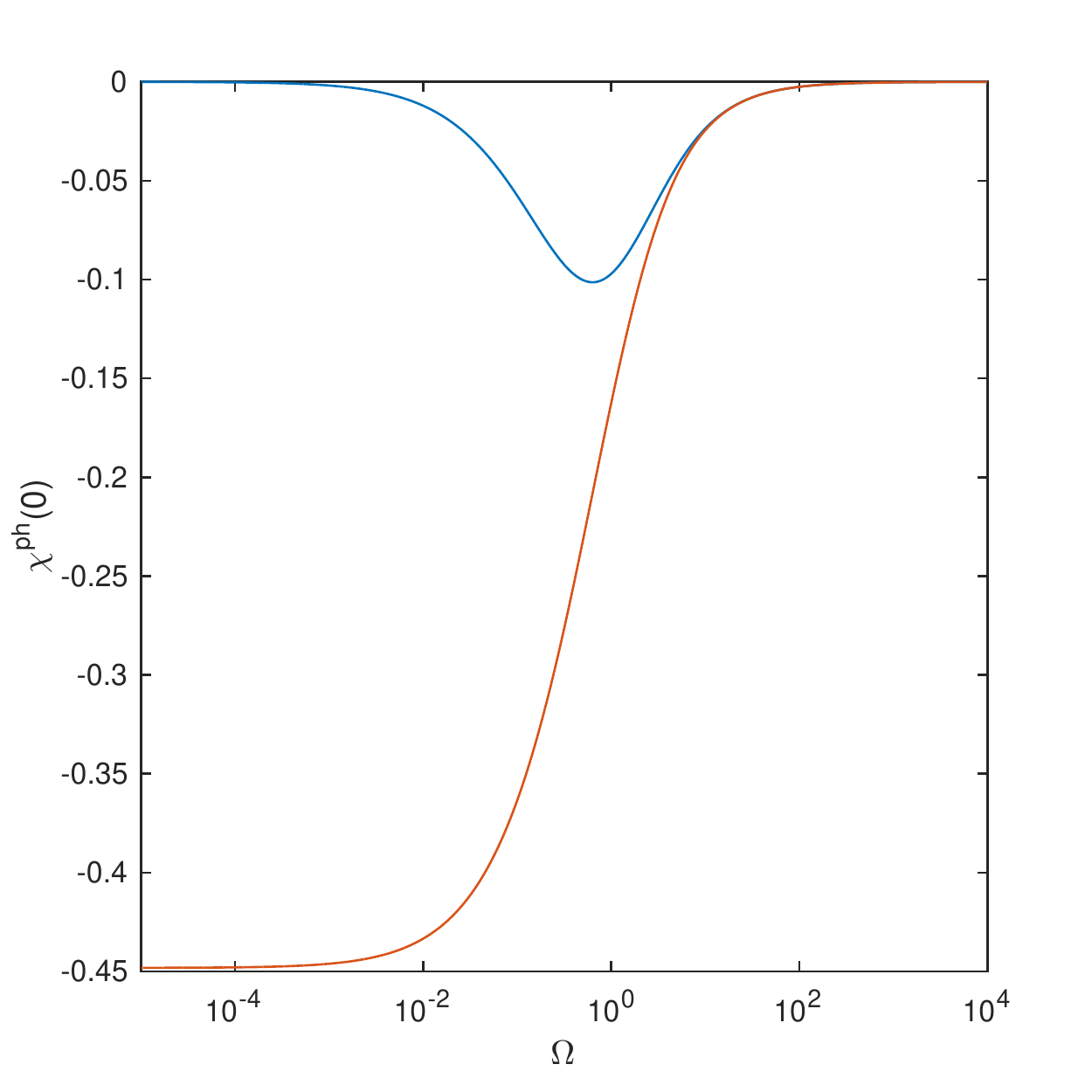}
\caption{Particle-hole bubble at zero momentum transfer vs.~$\Omega$, all in units of $t$, with intra-band (blue) and inter-band (red) components.}
\label{fig:phbub}
\end{figure}

The price to pay is some loss of unbiasedness, since with finite precision and a truncated form-factor basis, different assignments of the bare interaction onto the channels may produce different results. In particular, the Hubbard on-site term $U$ can be equivalently formulated as either density-density, spin-spin, or pairing interaction. One then introduces a slight bias towards magnetism when initializing $U$ fully as spin-spin interaction, for instance. Although splitting $U$ equally among the three channels is the most neutral choice, other possibilities can be considered for consistency checks, as is done in this work. As reference, the critical onsite coupling strength obtained by splitting $U$ equally among the three channels is $U_C = 2.7 \, t$, in accordance with the resulting $U_C$ when keeping $V^{\Omega_0}$ unassigned to any channel. Initializing $U$ in the magnetic or charge channels yields $U_C = 2.5 \, t$ and $3.1 \, t$, respectively. 

\section{Coulomb unscreening}\label{app:unscr}

%
\begin{figure}[t!]
\includegraphics[width=1.0\columnwidth]{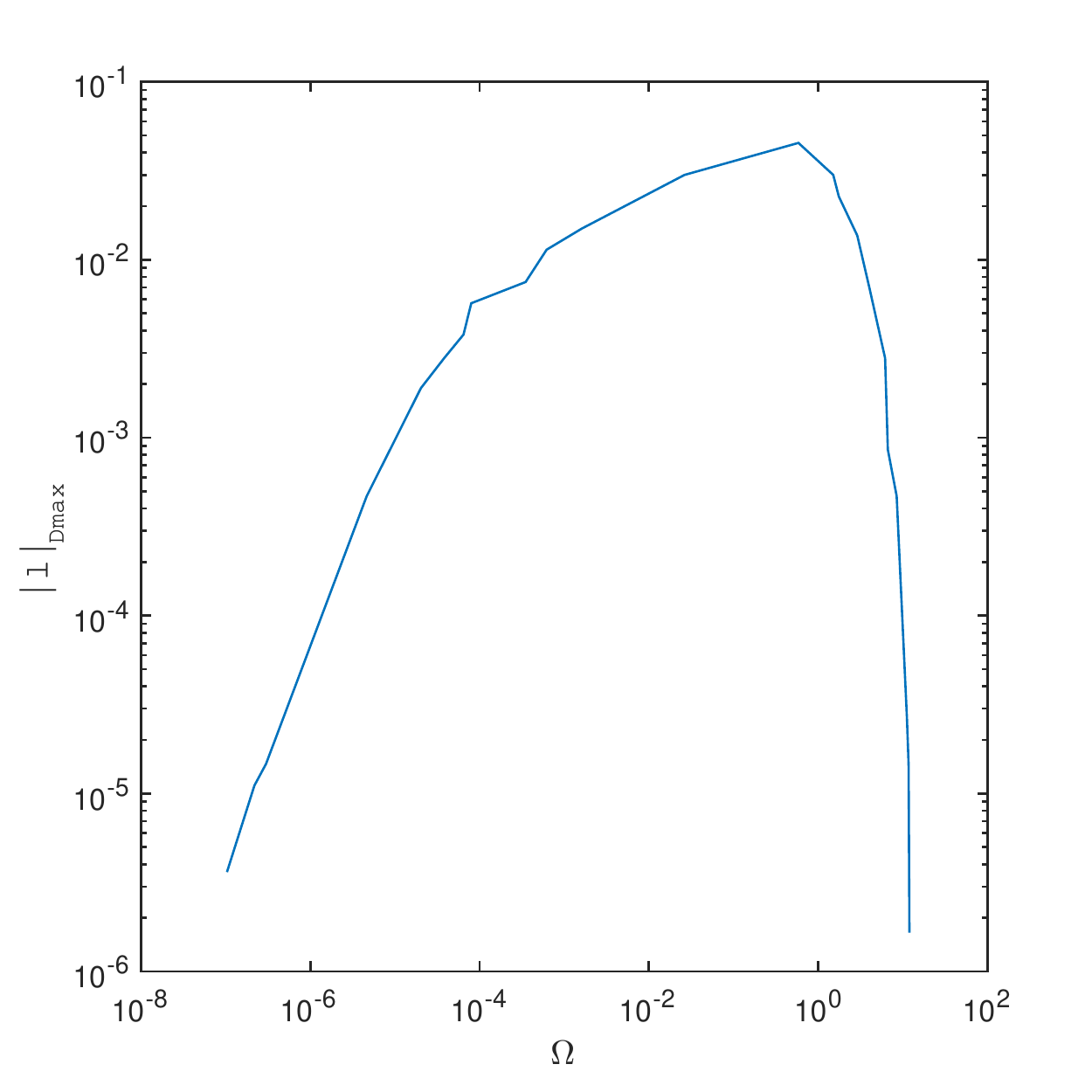}
\caption{Norm of transfer momentum $\bf{l}$ for which $D$ is maximal vs $\Omega$, all in units of $t$.}
\label{fig:Dpeak}
\end{figure}

Here we discuss the behavior of charge interactions at small $\Omega$ scales, where they experience anti-screening and flow back towards the unscreened bare Coulomb interaction due to the vanishing DOS. This can be understood by looking at the scale behavior of the particle-hole bubble $\chi^{\text{ph}}$, defined in Eq.~\eqref{eq:chis} and plotted in Fig.~\ref{fig:phbub}. The vanishing DOS suppresses the intra-band components as one approaches the Fermi level, resulting in a sign change of the $\Omega$-differentiated bubbles involved in the flow equations, cf. Eqs.~\eqref{eq:TUfRG_flow}, reversing the screening of electric charge. As already mentioned in the body text, this unscreening is not easy to deal with numerically. Any slight underestimation or overestimation of unscreening effects is magnified during the flow, resulting in either some screening persisting, or in charge interactions becoming fully unscreened before reaching the Fermi level. The overestimation is more problematic in practice, since Coulomb interactions suddenly grow huge and may even cause numerical overflows. Introducing a small chemical potential has no effect if it is smaller than the lowest scales we can flow to under this unscreening problematic. A larger $\mu$ of the order of such scales ($\sim 10^{-3}-10^{-2}$) naturally leads to a saturation of unscreening behavior, and as it corresponds to a system with finite DOS, screening remains. 

The unscreening problematic is also exacerbated by increasing the order of the ODE's solver, being more prone to overestimation and even displaying oscillating behavior unless the steps in $\Omega$ are taken to be unfeasibly small. Going over to a predictor-corrector scheme like the Adams-Basforth-Moulton multistep method cures the oscillations. 

Fig.~\ref{fig:Dpeak} illustrates another subtlety of the charge unscreening in the $\Omega$-regulator scheme. In the reconstruction of the bare Coulomb interaction taking place at low scales, the maxima of both particle-hole bubble and $D$ propagator do not stay at the $\Gamma$ point, but at small wave-vectors, which nonetheless tend towards $\Gamma$ as $\Omega \rightarrow 0$. The maxima of $\chi^{\text{ph}}$ start off at the K points for very high $\Omega$, wander inwards in the BZ and outwards again to the M points as one sweeps across the van Hove singularities in the flow, and inwards again towards the $\Gamma$ point. The $D$ propagator is peaked at $\Gamma$ almost for the whole screening stage. However, as the unscreening stage gets closer, both get peaked at small but finite momenta. The bubble peak is located at a bigger wavevector than the propagator, but follows the same trend as depicted in Fig.~\ref{fig:Dpeak}.

\bibliography{References}

\end{document}